\documentclass[preprint,aps]{revtex4}
\usepackage{graphicx}
\usepackage{dcolumn}
\usepackage{bm}
\usepackage{psfrag,amssymb}

\def\ni{\noindent}

\begin{document}
\sf

\title{ Frozen Fronts Selection in flow against self-sustained chemical waves}

\author{T. Chevalier}
\author{D. Salin}
\author{L. Talon}

\affiliation{Laboratoire Fluides Automatique et Syst\`{e}mes Thermiques,
Universit\'{e} Paris Sud, C.N.R.S. (UMR7608),
B\^{a}timent 502, Campus Universitaire, 91405 Orsay Cedex,
France}

\date{\today}

\begin{abstract}

Autocatalytic reaction fronts between two reacting species in the absence of
fluid flow, propagate as solitary waves. The coupling between autocatalytic reaction front and forced hydrodynamic flow may lead to stationary front whose velocity and shape depend on the underlying flow field. We focus on the issue of the chemo-hydrodynamic coupling between forced advection opposed to self-sustained chemical waves which can lead to static stationary fronts, i.e Frozen Fronts, $FF$.
Towards that purpose, we perform experiments, analytical computations and numerical simulations with the autocatalytic Iodate Arsenious Acid reaction ($IAA$) over a wide range of flow velocities around a solid disk. For the same set of control parameters, we observe two types of frozen fronts: an upstream $FF$ which avoid the solid disk and a downstream $FF$ with two symmetric branches emerging from the solid disk surface. We delineate the range over which we do observe these Frozen Fronts. We also  address the relevance of the so-called eikonal, thin front limit to describe the observed fronts and select the frozen front shapes.
\end{abstract}

\maketitle

\vspace{1in}


\section*{Introduction}
 Depending on the reaction kinetics, chemical reaction fronts exhibit fascinating phenomena such as Turing patterns, Belousov-Zhabotinsky oscillations, and chaotic or solitary wave propagation \cite{scott94}. Autocatalytic reactions lead to fronts propagating as solitary waves with a constant velocity and invariant, flat, concentration profile resulting from a balance between reaction and diffusion \cite{scott94,fisher37,kolmogorov37}. These fronts are analogous to flames in combustion \cite{zeldovich38} and autocatalytic reactions are a kind of "cold combustion model" especially in the thin flame limit. In contrast to flame propagation in combustion \cite{zeldovich38}, where it has been analyzed thoroughly
theoretically and experimentally, the effect of fluid flow (laminar or turbulent) on reaction fronts has not been explored in detail until recently \cite{audoly00,edwards02,edwards06,leconte03,leconte04,vasquez07,leconte08,schwartz08,
mitchell12,bargteil12,megson15,mahoney15,atis12,saha13}. In the presence of an hydrodynamic flow, it has already been observed and understood that such fronts while propagating at a new constant velocity, adapt their shape in order to achieve a balance between reaction diffusion and flow advection. More recently, the focus has been on the situation where the flow field acted  against the chemical reaction. In such a case, it has been observed over a wide range of flow velocity that the fronts are neither propagating  forward (in the chemical reaction direction) nor blown in the flow direction but remains static, frozen. In this dynamical equilibrium, chemistry and flow are both at work. For instance in porous media \cite{saha13}, the front is pinned around the stagnation zones of the flow, due to the porous structure, and the front is distorted, curved in order to accommodate the local flow velocity fluctuations. In cellular flows,the {\it frozen fronts} are pinned in vortex structure \cite{mitchell12,mahoney12}.\\

To visualize the frozen fronts ($FF$) , we designed an experiment with the Iodate Arsenous Acid ($IAA$) chemical reaction in a simple heterogeneous forced flow field, namely a constant flow around a single disk-obstacle, opposed to the natural autocatalytic reaction front propagation. Depending on the control parameters we do observe two types of $FF$
: one upstream $FFs$  which avoid the solid disk and one downstream $FFs$ with two symmetric branches emerging from the solid disk surface. We delineate the range over which we do observe these Frozen Fronts. Numerical simulations provide a systematic phase diagram of the Frozen Fronts. Using the so-called eikonal limit of thin front thickness, we are able to account for the selection rule of the Frozen Fronts.

\section{Experiments}
\begin{figure}[htbt]
\begin{center}
\includegraphics[width=6cm]{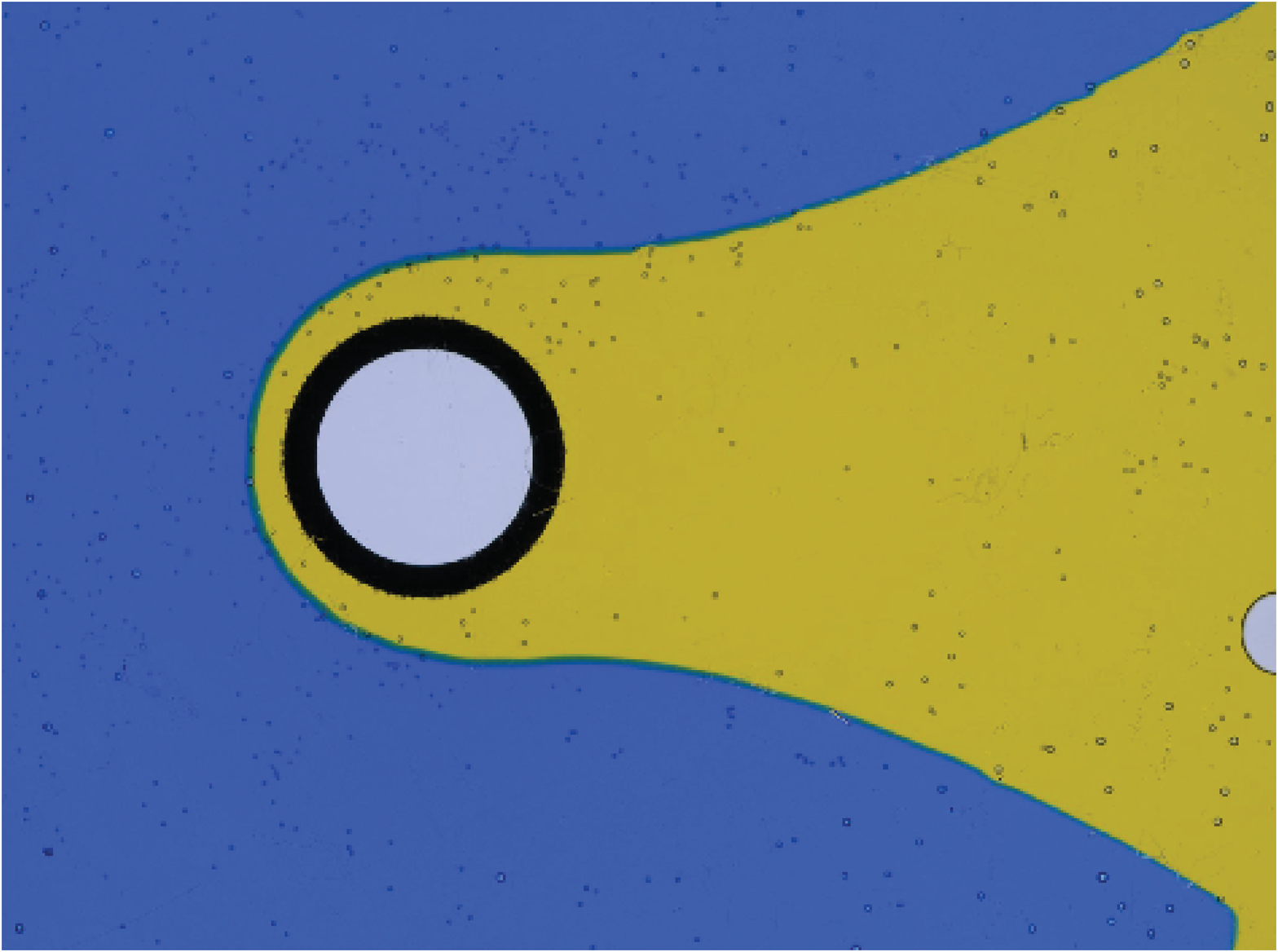}
\includegraphics[width=6cm]{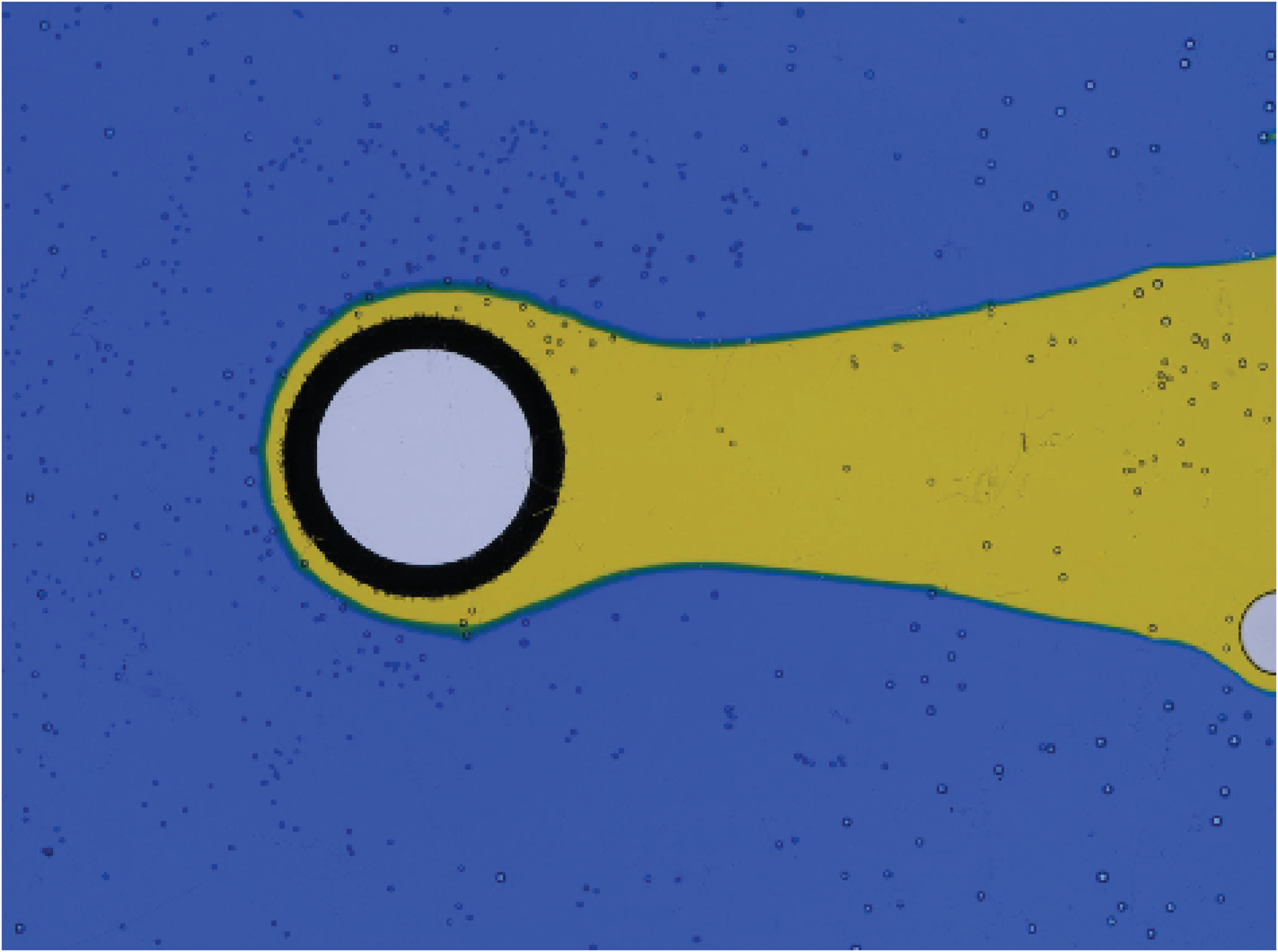}
\includegraphics[width=6cm]{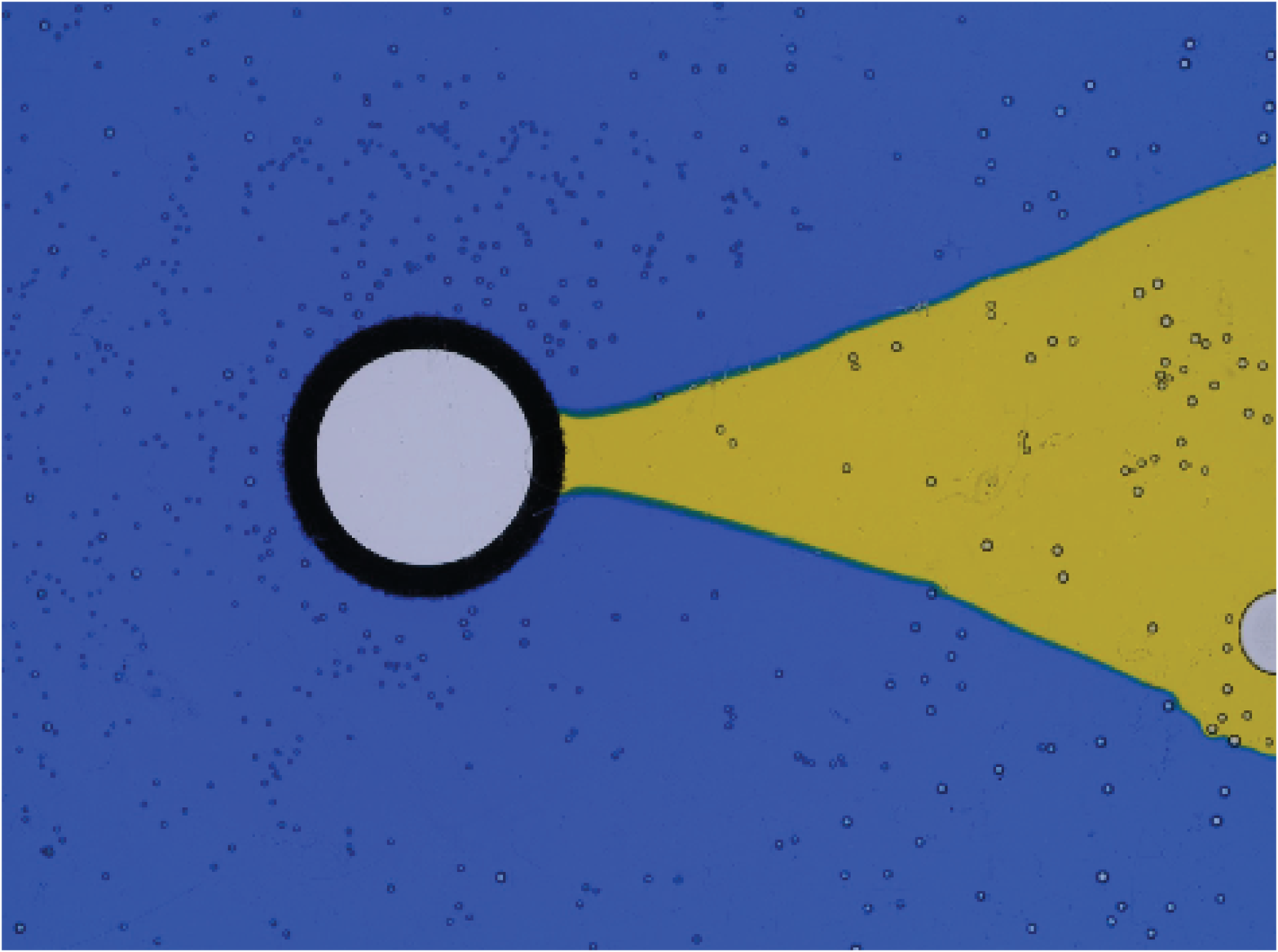}
\includegraphics[width=6cm]{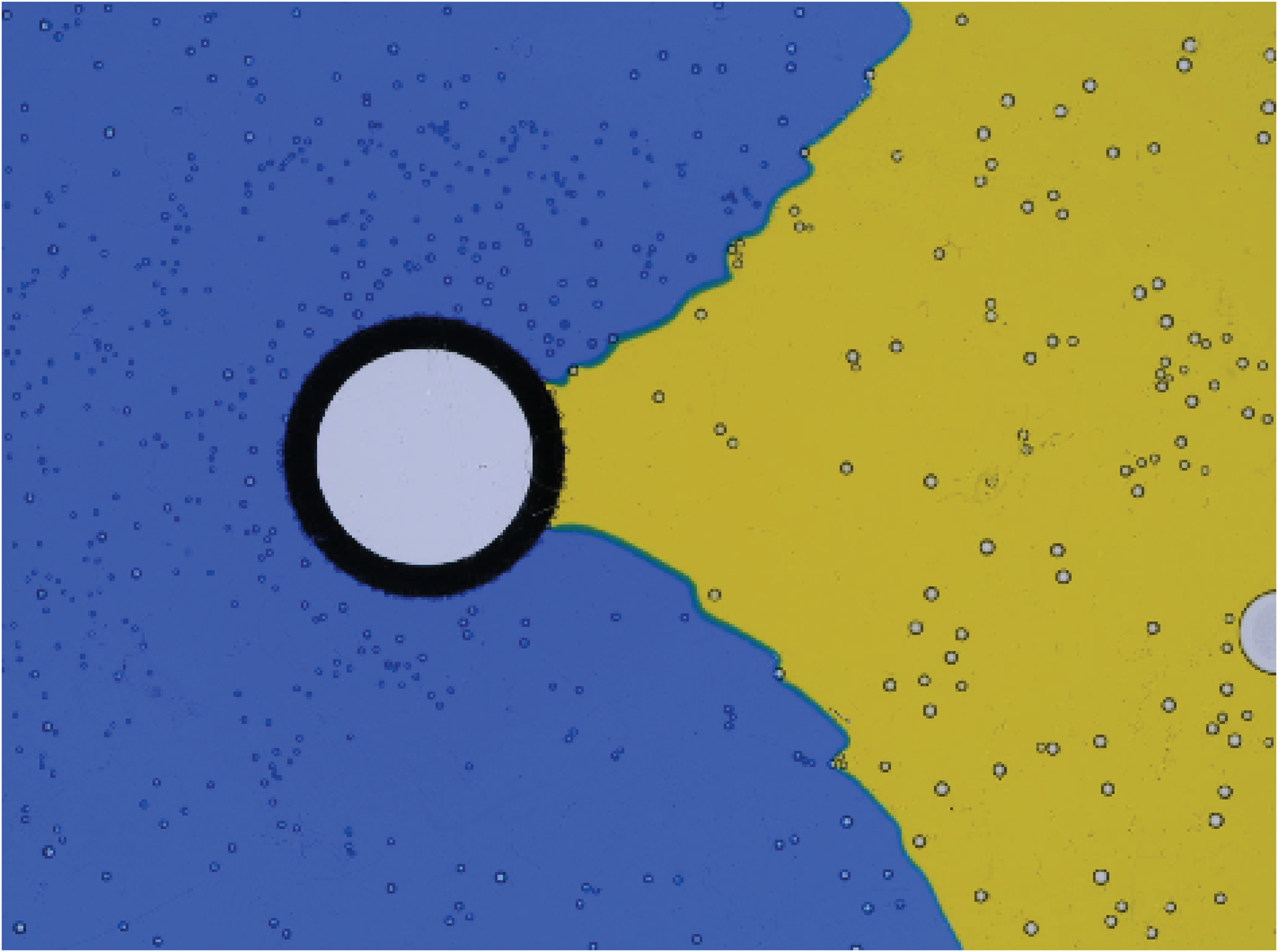}
\caption{\small  Frozen chemical front in a flow around a circular disk in a Hele-Shaw cell. The fresh, blue reactant is injected from left to right at a flow at constant velocity $U_0$. The solid disk obstacle is a cylinder of diameter $17 mm$. The chemical front in the absence of flow would propagate from right to left at a velocity $V_{\chi}$. Left pictures : the two frozen fronts observable for the same flow velocity $U_0/V_{\chi}=-2.5$: one type of front corresponds to an upstream front that avoids the solid, the other type corresponds to a downstream front originating from the solid disk. Top right : the upstream $FF$ front for $U_0/V_{\chi}=-5$. Bottom right the downstream $FF$ front for $U_0/V_{\chi}=-1.4$.}
\label{solid}
\end{center}
\end{figure}

We performed experiments with the Iodate Arsenous Acid ($IAA$)
autocatalytic reaction:
\begin{equation}\label{eqchimie}
 3 \mbox H_{3}\mbox{AsO}_{3} + \mbox{IO}_{3}^- + 5
 \mbox I^- \longrightarrow 3 \mbox H_{3}\mbox{AsO}_{4} + 6 \mbox I^-
\end{equation}
The reaction is autocatalytic in iodide (I$^{-}$). The
concentrations used are: $[\mbox{IO}_{3}^{-}]_{0}=7.5$ mM,
$[\mbox H_{3}\mbox{AsO}_{3}]_{0}=25$ mM. As the ratio,
$[\mbox H_{3}\mbox{AsO}_{3}]_{0}/[\mbox{IO}_{3}^{-}]_{0}>3$, the Arsenous is in excess \cite{hanna82} and the front can be localized by the transient iodine generated during the reaction. Instead of the usual method using starch to detect the transient
iodine, we use Polyvinyl alcohol ($PVA$) at a concentration of $6$ kg/m$^3$ which is much more sensitive \cite{yoshinaga04}  and also gives a good optical contrast (Figs. \ref{dipole} and \ref{bubble}). In addition, we add to the fluids bromocresol green $PH$ sensitive dye which gives the position of the leading edge of the reaction front: its color is blue for reactants and yellow for products. In some experiments we only use the later front detection (Fig. \ref{solid}).

This autocatalytic reaction in the fluid flow of local velocity $\vec{U}$ is govern by the convection(or advection)-reaction-diffusion equation which can be written as:
\begin{equation}
\frac{\partial C}{\partial t}+\overrightarrow{U} \overrightarrow{\nabla}C=D_{m}\triangle
C+\frac{1}{\tau}f(C)
\label{CRD}
\end{equation}
where the specific kinetics of the $IAA$ reaction is third order \cite{hanna82}: $f(C)=C^{2}(1-C)$. $C$ is the concentration of the autocatalytic
reactant (iodide), normalized by the initial concentration of
iodate ($C=[\mbox I^{-}]/[\mbox{IO}_{3}^{-}]_{0}$), $\tau$ the reaction time and $D_m$ the molecular diffusion. In the absence of flow, $U=0 \; m/s$, the balance
between diffusion and reaction leads to a solitary wave of
constant velocity $V_{\chi}$ and width $l_{\chi}$ 
\cite{scott94,hanna82,bockmann00}, solutions of Eq. (\ref{CRD}) given by:
\begin{equation}
  V_\chi=\sqrt{\frac{D_m}{2 \tau}} \, , \hspace{1cm}
   l_\chi=D_m /V_\chi \, , \hspace{1cm} C=\left(1+\exp-(\frac{x-V_{\chi} t}{l_{\chi}})\right)^{-1}
  \label{eq:vl}
\end{equation}
where $x$ is the propagation direction of the wave.
With the above concentration, we measure $V_{\chi}=(11 \pm 1)\, \mu$m/s, from which we can infer the reaction front width, $l_{\chi}\simeq 200 \; \mu m$.
To achieve a quasi-$2D$ velocity field we use a Hele-Shaw cell  \cite{Heleshaw98}, two thick rigid transparent parallel plates separated by a small gap ($250 \; \mu m$). The solid circular disk of diameter $2R=17 mm$ is a joint quenched between the two plates. The uniform flow rates is achieved by two pumps, one for each inlet (far upstream from the disk), and three outlets. The $2D$ flow field around the disk in a uniform far field in a Hele-Shaw is well known: at leading order it is the potential flow of a uniform flow around an hydrodynamic dipole \cite{lamb32} which intensity is linked to $R$.
For a viscous flow the velocity must be zero on the boundary, therefore there are boundary layer type corrections which extend over a typical size of the order of the gap thickness which is analytically fitted \cite{lee69}.
It is worth noting already that we have also performed experiments around an hydrodynamic dipole where we inject and suck product at the same flow rate between two inlet outlet distant of $5 \; mm$ (see Fig. \ref{dipole}) and around an injected air bubble (Fig \ref{bubble}).

\section{Two types of frozen front in uniform flow past a solid disk}
We have performed a series of experiments for different flow rate $U_0$ opposed to the chemical wave without flow. Let us call $u=|U_0/V_{\chi}|$ the control parameter of the experiment. In the experimental procedure, we first initiate the front and let it propagate for a while; then at chosen time we switch on the two pumps to generate an uniform flow. For $u<1$, fronts propagate always to the left and never stop, there is no frozen front. For $u>1$, after a transient,  Frozen, static fronts built on :  Fig. \ref{solid} is a plot of such frozen fronts. For the same $u$, we did observe two different types of Frozen fronts depending on the switching time: for initial front generated on the left of the disk, the "upstream" $FF$ keeps to avoid touching the disk surface. When the initial front is in contact with the disk, the front keep this contact later on leading to a "downstream" $FF$ with two symmetric branches. In Fig. \ref{solid} we see that increasing $u$ result for the upstream front to be closer to the solid , whereas for the downstream one, the two branches of the front are closer to the symmetry axis.\\
For  both front types, for $u\succeq 5-6$, the front can not keep a stationary shape and hence no frozen fronts are observed. For upstream front, the forming fronts comes so close to the left of the solid disk that it comes into contact with it and then is transported by the flow to the right. For downstream fronts, the separation between the two branches becomes thinner and thinner as $u$ increases and can lead to a pinch-off, followed by a detachment as observed in \cite{atis12}.

\section{Frozen front selection using the Eikonal equation}

As $l_{\chi}\simeq 200 \; \mu m$ is rather small compared to the disk size ($2R=17 \;mm$) and as the experimental front is reduced to a single iso-concentration (in some instance we have been able to get two iso-concentrations as in Figs. \ref{dipole} and \ref{bubble}), it is tempting to describe this constant concentration line in the framework of the eikonal thin front approximation.
When the front width $l_\chi$ is much smaller than the typical size of the system (disk radius) the so called eikonal account accurately of the front behavior \cite{edwards02,leconte04}. It also corresponds to the thin flame regime in combustion. In such regime  Eq. \ref{CRD} is replaced by the front evolution :
\begin{equation} \label{eikonalpropa}
\frac{d\overrightarrow{r}}{dt}=\vec{V_{F}}=\vec{U}+V_{\chi} (1 + l_{\chi} \kappa)\cdot\vec{n}
\end{equation}

\begin{equation} \label{eikonal}
\vec{V_{F}}\cdot\vec{n}=\vec{U}\cdot\vec{n}+V_{\chi} + D_m \kappa
\end{equation}

where $\vec{U}$ is the local fluid velocity at the front position $\overrightarrow{r}$, $\vec{n}$ is the local unit vector normal to the interface (oriented from product to reactant) and $\kappa$ is the curvature of the interface.

\begin{figure}[htbt]
\begin{center}
\includegraphics[width=6cm]{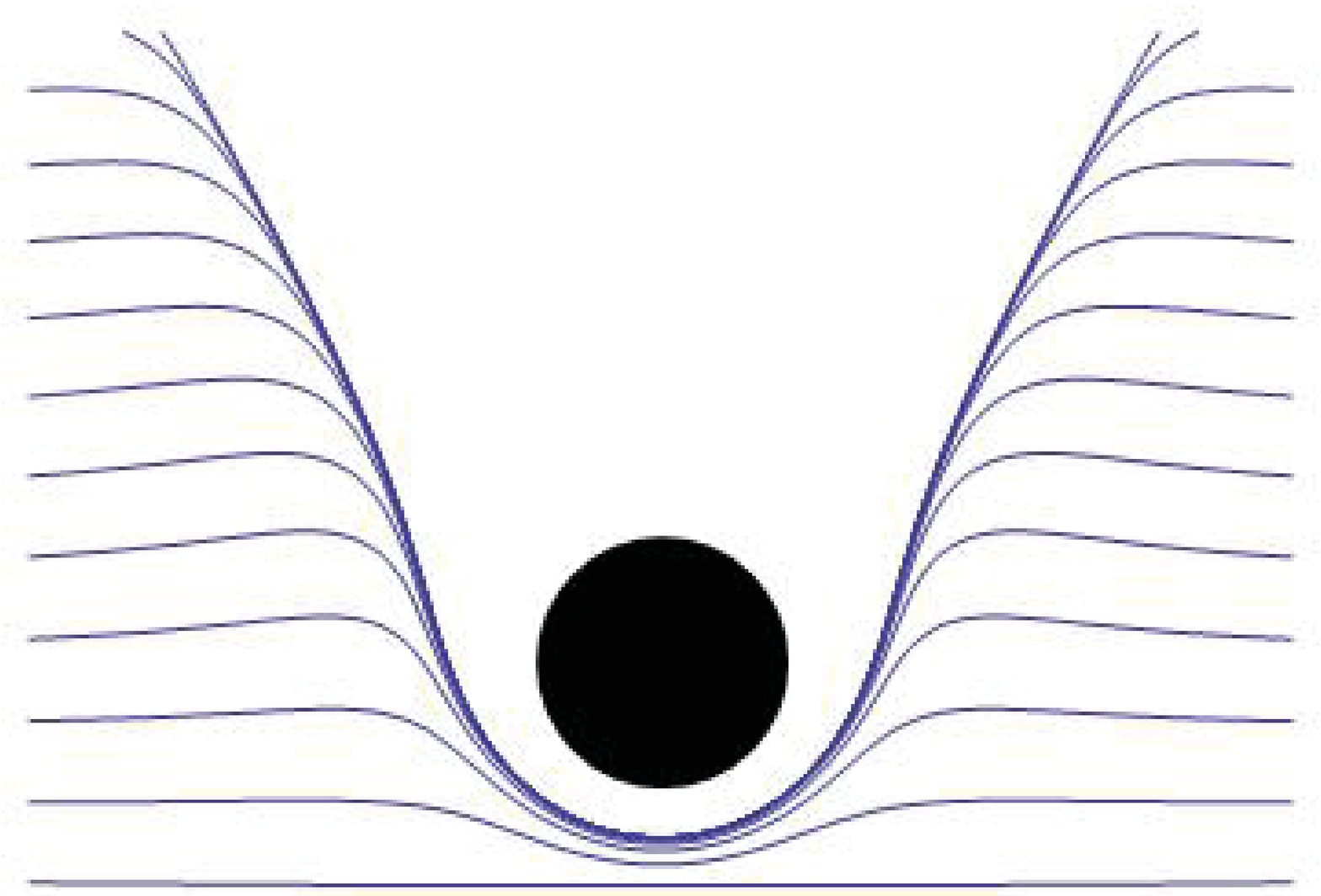}
\includegraphics[width=6cm]{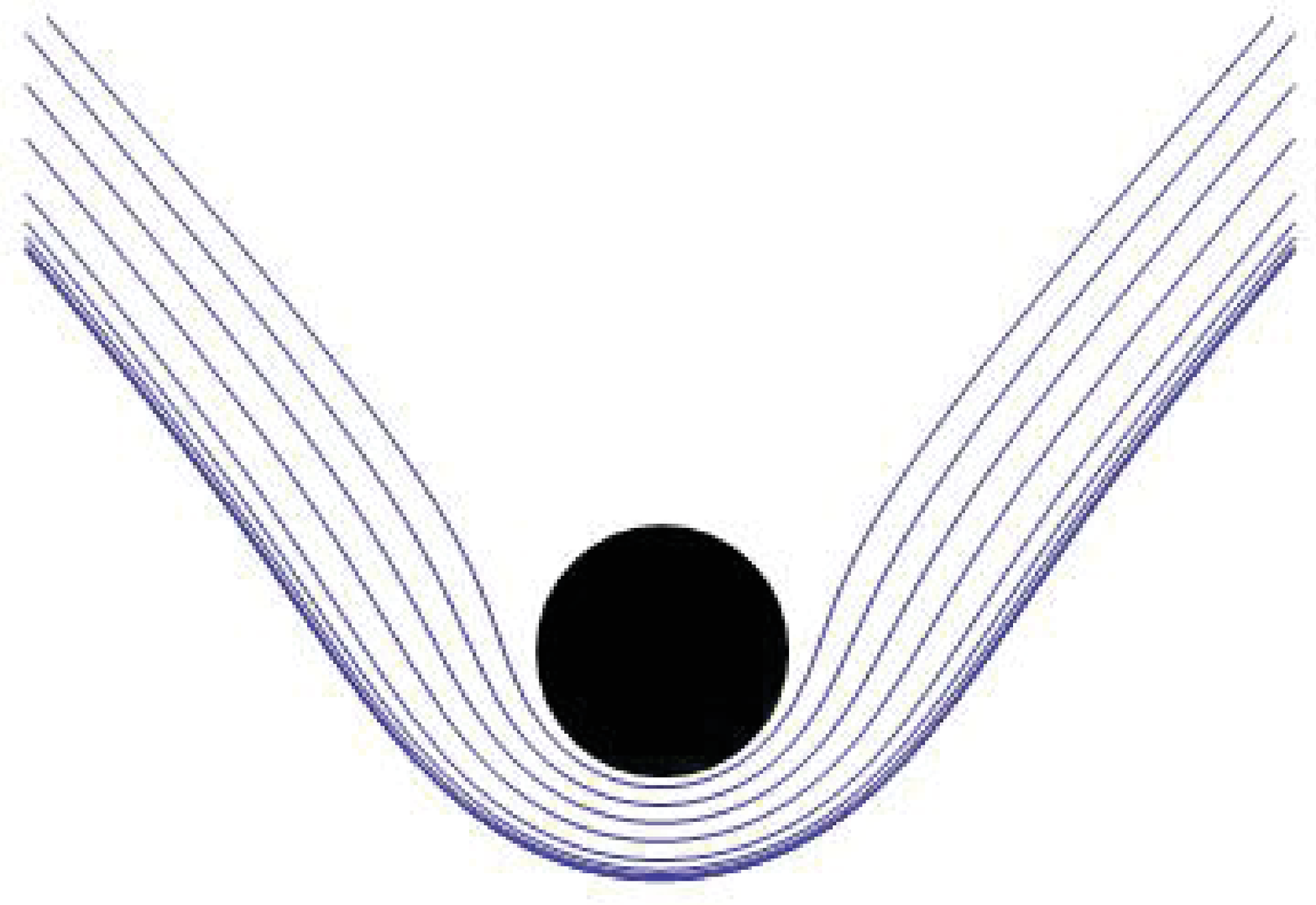}
\includegraphics[width=8cm]{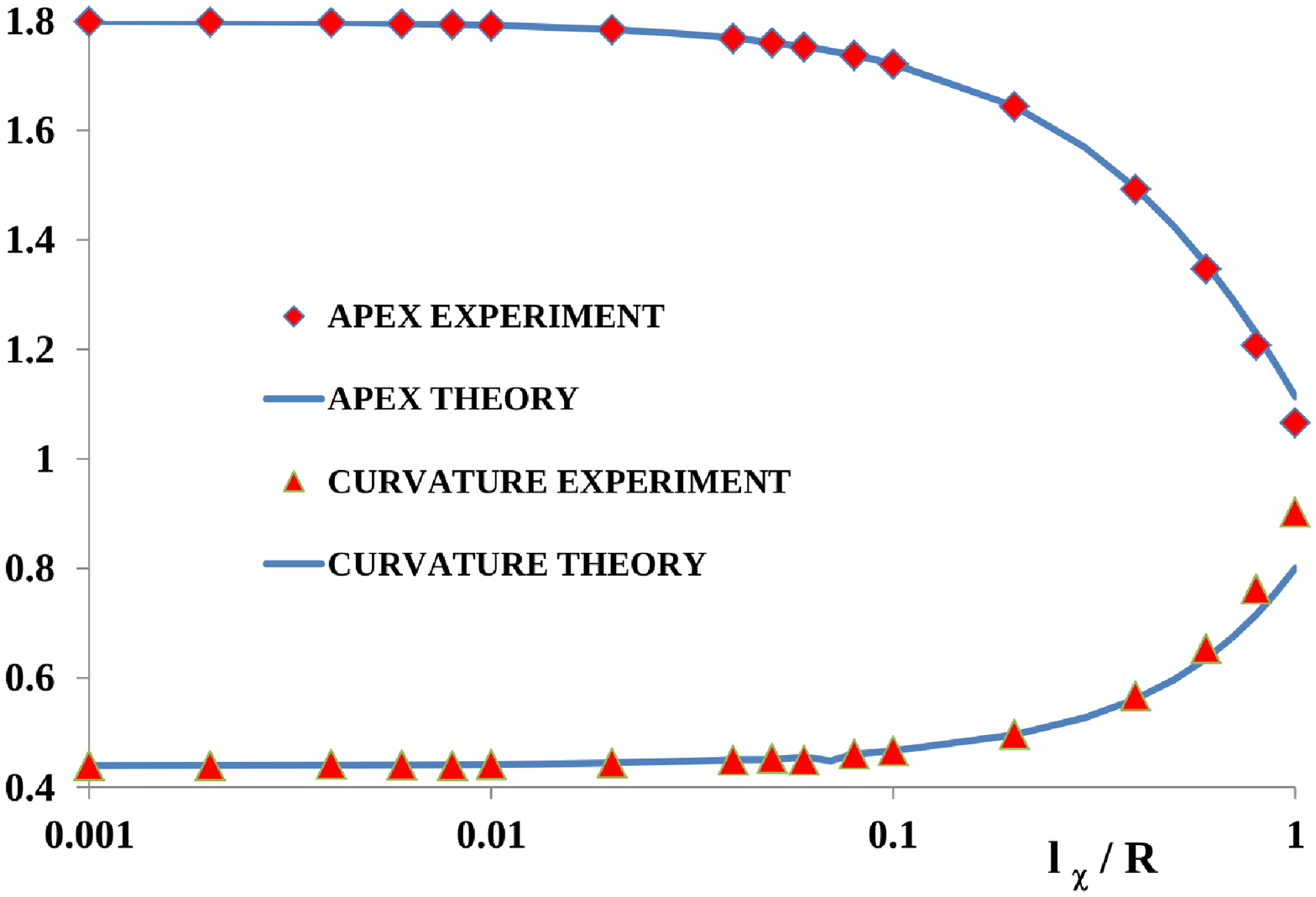}
\includegraphics[width=8cm]{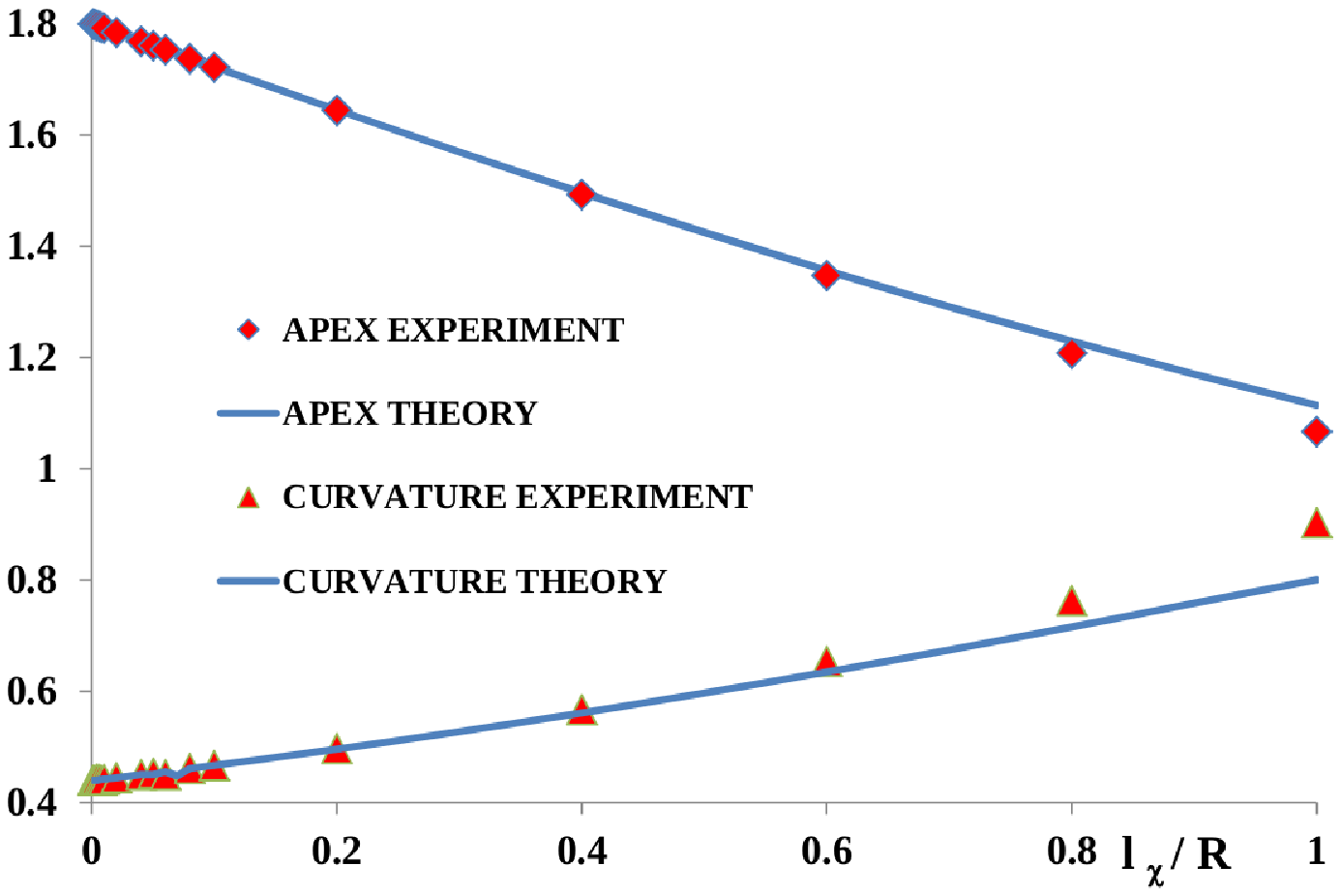}
\caption{\small  Frozen chemical front in a flow around a disk using the eikonal equation \ref{eikonalpropa}. Top left : dynamics of the frozen front formation, front at different time steps ($u=2$, $l_{\chi}/R=0.1$). Top right : achieved static frozen front for different values of $l_{\chi}$ ($u=1.5$). Bottom : plot of the distance $|h(0)|$  from the disk (on the symmetry axis) and the corresponding curvature versus the $l_{\chi}/R$; Log-linear and linear plots (inserted) :  The squares and triangles correspond respectively to the distance $|h(0)|$ (top) and the curvature, $h''(0)$ (bottom). The solid curves corresponds to the theoretical selection (see text).}
\label{selection}
\end{center}
\end{figure}

As $l_{\chi}$ is assumed small, the eikonal equation is usually used neglecting the last curvature term \cite{mahoney15}. Indeed this assumption is a little inconsistent as it discard the molecular diffusion which is part of the reaction diffusion process (see Eq. \ref{eq:vl}). Moreover from the mathematical point of view, it let the front curvature undetermined as it will be discussed later on.

We will use for the velocity field the $2D$ velocity field around a solid obstacle that will be used in the following numerical simulations. We also address in this section only the class of upstream fronts that avoid the solid obstacle.

The top left figure in Fig. \ref{selection} shows the dynamics of the formation of the frozen front at different time steps for $u=1.5$ and $l_{\chi}/R=0.1$. This dynamic is obtained by a step by step direct integration of the full Eq. \ref{eikonalpropa} with a flat front far away upstream the disk as the initial condition. Obviously, the front achieved an asymptotic Frozen Front. The top right of the same figure shows the asymptotic frozen front, $h(y)$ (where $y$ is perpendicular to the symmetry axis, $x$), for different values of $l_{\chi}$: the selected frozen front definitively depends also on the chemical length $l_{\chi}$.The larger $l_{\chi}$, the smaller the distance (apex) of the front, $|h(0)|$, from the disk. The corresponding data of $h_0$ versus $l_{\chi}$ are the squares in the bottom of Fig. \ref{selection}. On the same figure we give the corresponding curvature at the apex, $h''(0)$ (triangles). For a given set of parameters ($u$,$l_{\chi}/R$) the asymptotic $FFs$ obtained from the dynamics Eq. \ref{eikonalpropa} have a characteristic shape which can be characterized by the position on the axis $h(0)$ and the curvature $h''(0)$ for instance. How can we account for the observed shape of the frozen front? The frozen front is time independent
 ($\overrightarrow{V_F}=\overrightarrow{0}$) and hence should follow the static eikonal equation,
$\vec{U}\cdot\vec{n}+V_{\chi} (1 + l_{\chi} \kappa)=0$ which writes:

 \begin{equation} \label{eikonalxy}
  -\frac{v_x(h(y),y)}{\sqrt{h'(y)^2+1}}+\frac
   {h'(y)
   v_y(h(y),y)}{\sqrt{h'(y)^2+1}}+1-\frac{l_{\chi} h''(y)}{\left(h'(y)^2+1\right)^{3/2}}=0
\end{equation}
where $v_x(x,y)$ and $v_y(x,y)$ are the $x$ (along the symmetry axis in the $\overrightarrow{U_0}$ direction) and $y$ (transverse direction) components of the dimensionless velocity field $\overrightarrow{v}=\overrightarrow{U}/V_{\chi}$. $h(y)$ is the eikonal front. With this chosen axis orientation, in Figs \ref{solid} and  \ref{selection} the curvature $h''(0)$ is positive and hence reduces the effective chemical velocity on the symmetry axis by a factor: $1-l_{\chi} h''(0)/\left(h'(0)^2+1\right)^{3/2}$.

Eq. \ref{eikonalxy} is a second order differential equation for $h(y)$ which integration requires two conditions. Lets us start the integration from the point on the $x$ axis, $(h(0),0)$. Owing to the symmetry of the problem, we have $h'(0)=0$. Inserting the later in Eq. \ref{eikonalxy} leads to $-v_x(h(0),0)+1-l_{\chi} h''(0)=0$ : for each curvature $h^{''}(0)$ on the symmetry axis, there is at least one value of  $h(0)$ that fulfill Eq. \ref{eikonalxy}, therefore there is no obvious selection of the Frozen Front. Note that if we take the $h(0)$ value obtained from the asymptotic frozen front obtained from the dynamics, we do recover the same frozen front integrating Eq. \ref{eikonalxy} with the two conditions ($h(0)$,$h'(0)$). To address the issue of the selection of $h(0)$ let us make the Taylor expansion of the static eikonal equation in the vicinity of the apex of the frozen front ($y\simeq 0$) for the variables involved in Eq. \ref{eikonalxy} namely
\begin{eqnarray}\label{dl}
h(y)=h(0)+\frac{1}{2} y^2  h''(0) +\frac{1}{24} y^4 h^{4}(0)+ O(y^{6}) \\
v_x(h(y),y)=v_x(h(0),0)+\frac{1}{2} y^2 \left(h''(0)
   v_x(h(0),0)+\partial_{xx}v_x(h(0),0)\right)+O(y^{4})\\
v_y(h(y),y)=-y
  \partial_{x}v_x(h(0),0)-\frac{1}{6} y^3 \left(3 h''(0)
   \partial_{xx}v_x(h(0),0)+\partial_{xyy}v_x(h(0),0)\right)+O(y^{4})
\end{eqnarray}
where we have taken into account the symmetries on the x axis, namely, $h(y)=h(-y)$, $v_x(h(y),y)=v_x(h(-y),-y)$ and  $v_y(h(y),y)=-v_y(h(-y),-y)$ and of the fluid incompressibility ($\partial_{x}v_x+\partial_{y}v_y=0$). Injecting these expansions in Eq. \ref{eikonalxy} leads to :
\begin{eqnarray}\label{eikexpansion}
\nonumber
 0= -v_x(h_0,0)+1-l_{\chi}h''_0+\\
\nonumber
y^{2}\{v_x(h_0,0)h''^2_0-3\partial_{x}v_x(h_0,0)h''_0
  -\partial_{yy}v_x(h_0,0)+ l_{\chi}(3h''^3_0-h_0^{(4)})\}+\\
\nonumber
\frac{y^{4}}{24} \{6 h''^2_0 \partial_{yy}v_x(h_0,0)-15 h''^2_0
   \partial_{xx}v_x(h_0,0)+2 h''_0
   \partial_{xyy}v_x(h_0,0)-(5 h^{(4)}_0+6 h''^3_0)+\\
\nonumber
   \partial_{x}v_x(h_0,0)+(4 h^{(4)}_0 h''_0-9h''^4_0)
   v_x(h(0),0)-\partial_{yyyy}v_x(h_0,0)-l_{\chi}
   (h^{(6)}_0-30 h^{(4)}_0  h''^2_0+45
   h''^5_0))\}+\\
\nonumber
O(y^{6})
\end{eqnarray}
where for compactness we use $h_0=h(0)$, $h''_0=h''(0)$ etc.
The leading order ($y^{0}$) in the above relationship involves $h(0)$ and $h''(0)$ . The second order ($y^{2}$) involved $h(0)$, $h^{''}(0)$ and $h^{4}(0)$. With such an expansion, solving Eq. \ref{eikonalxy} can be achieved by equating to zero the coefficients of each order. We thus get a set of equations corresponding to each order. To be explicit, at order $O(y^{0})$ we get:
  \begin{equation} \label{eikonalzero}
-v_x(h_0,0)+1-l_{\chi} h''_0=0
\end{equation}
and at order  $O(y^{2})$:
\begin{equation} \label{eikonalsecond}
v_x(h_0,0)h''^2_0-3\partial_{x}v_x(h_0,0)h''_0
  -\partial_{yy}v_x(h_0,0)+ l_{\chi}(3h''^3_0-h_0^{(4)})=0
\end{equation}

Each order, $0(y^{2n})$ involves one derivative, $h_0^{2n+2}$ of the next order $0(y^{2n+2})$, therefore a method of solution consist of successive improving approximations: the leading order Eq. \ref{eikonalzero} discarding the last term $l_{\chi} h''_0$ leads to the zero order first approximation for $h_0$. Using  Eq \ref{eikonalzero} together with Eq. \ref{eikonalsecond}, discarding $h_0^{4}$, leads to a second order approximation for $h_0$ and $h''_0$. Using the zero, second and fourth order equations together discarding $h_0^{6}$ leads to a fourth order approximations for $h_0$, $h''_0$ and $h_0^{4}$ etc. We have checked that increasing the number of orders, the values obtained through this procedure converge to the same values. On the bottom of Fig. \ref{selection}, where the measured values of $|h_0|$ and $h''_0$  on the asymptotic frozen fronts are plotted versus $l_{\chi}/R$,  the solid curves correspond to the second order approximation; note that the fourth order is completely indistinguishable of these curves from the second one. The agreement is rather good, validating our procedure to account for the selection of the frozen fronts. Moreover the two plateaus observed as $l_{\chi}\rightarrow 0$ in the Log-linear plot at the bottom of the same Fig. \ref{selection} clearly shows that both  $h(0)$ and $h''(0)$ tend to finite values which are the full eikonal limit ones  $h(0)_{l_{\chi}\rightarrow 0}$ and $h''(0)_{l_{\chi}\rightarrow 0}$.

\begin{figure}[htbt]
\begin{center}
\includegraphics[width=6cm]{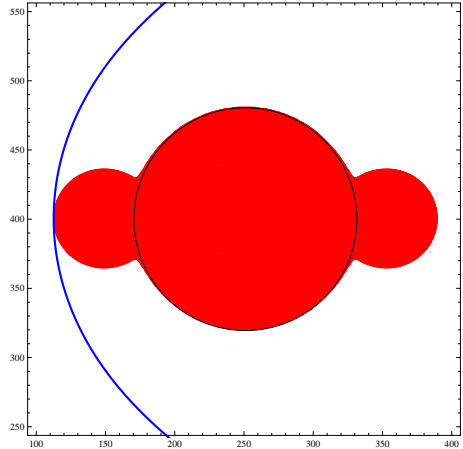}
\caption{\small The red zone around the solid disk (black circle) corresponds to the so-called slow zone where $|v|<1$. The solid blue line corresponds to the full eikonal frozen front ($l_\chi =0$).}
\label{Slowzone}
\end{center}
\end{figure}

This allows us to revisit the "classical" use of the eikonal equation \cite{mahoney15} which neglects the last term in Eq.\ref{eikonal} : indeed setting $l_{\chi}=0$ in Eq.\ref{eikonal} does not mean that there is no front curvature but that the leading zero order, $-v_x(h_0,0)+1=0$, gives only the front position, $h_{l_{\chi}=0}(0)$ which is the same as the one we get  $h(0)_{l_{\chi}\rightarrow 0}$. The second order, Eq. \ref{eikonalsecond}, with $h_{l_{\chi}=0}(0)$ and $l_{\chi}=0$ leads to $h''_{l_{\chi}=0}(0)$  etc.  Therefore our selection procedure gives at the second order the required curvature, leading to a complete selection of the frozen front which is, as verified, identical to the direct integration of $\vec{U}\cdot\vec{n}+V_{\chi}=0$. In the same papers \cite{mahoney15}, a clever construction of the allowed fronts is described. In the full eikonal limit ($l_{\chi}=0$), the front is initiated at the boundary of slow zones and is not allowed to penetrate them. These slow zones correspond to $|\overrightarrow{U}|\leq V_{\chi}$. In Fig. \ref{Slowzone}, we have drawn the slow zone around the solid disk (red) and the  full eikonal frozen front ($l_{\chi}=0$) corresponds to the blue solid line which obviously follow the prescription.

\section{Comparison with the experiments on uniform + dipole flow field}
We want to test these selection predictions against the experiments. Due to the contact with the solid disk mentioned above, the range of accessible $u$ values was limited. Therefore, we design an analogous experiment which allows a wider range of $u$ values. If we remember that the potential flow around a solid disk in a Hele-Shaw is nothing but the one around an hydrodynamic dipole (immaterial solid), we can design such an experiment. The fresh, blue reactant is injected from left to right as in Fig. \ref{solid}  at a flow at constant velocity $U_0$.
The "solid" obstacle is mimicked by an hydrodynamic dipole, injecting burnt, yellow product by the leftmost hole (source) and sucking it from the other hole (sink) at the same flow rate ($U_+=-U_-$); the distance between holes is $d=5 \;mm$.
\begin{figure}[htbt]
\begin{center}
\includegraphics[width=6cm]{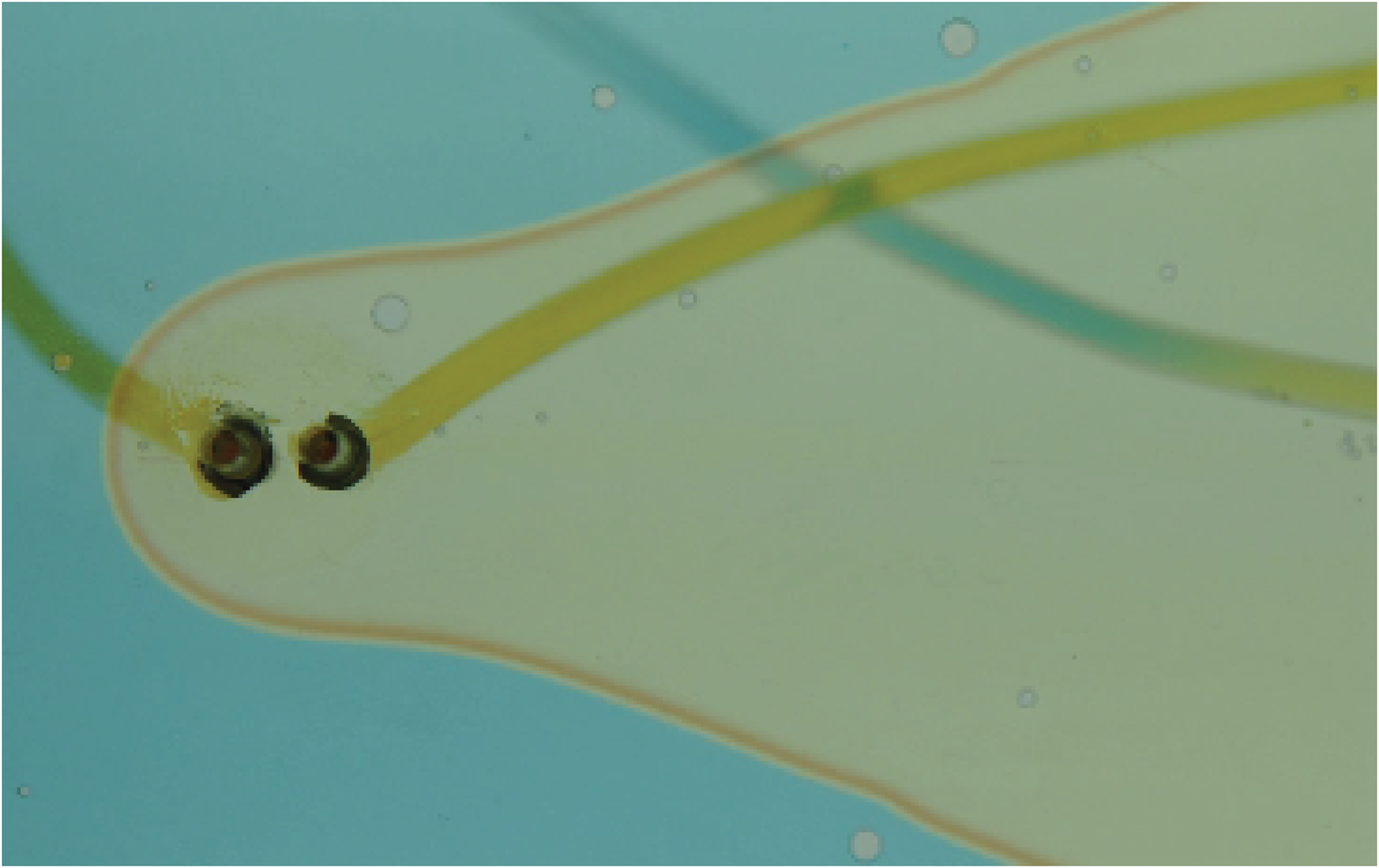}
\includegraphics[width=6cm]{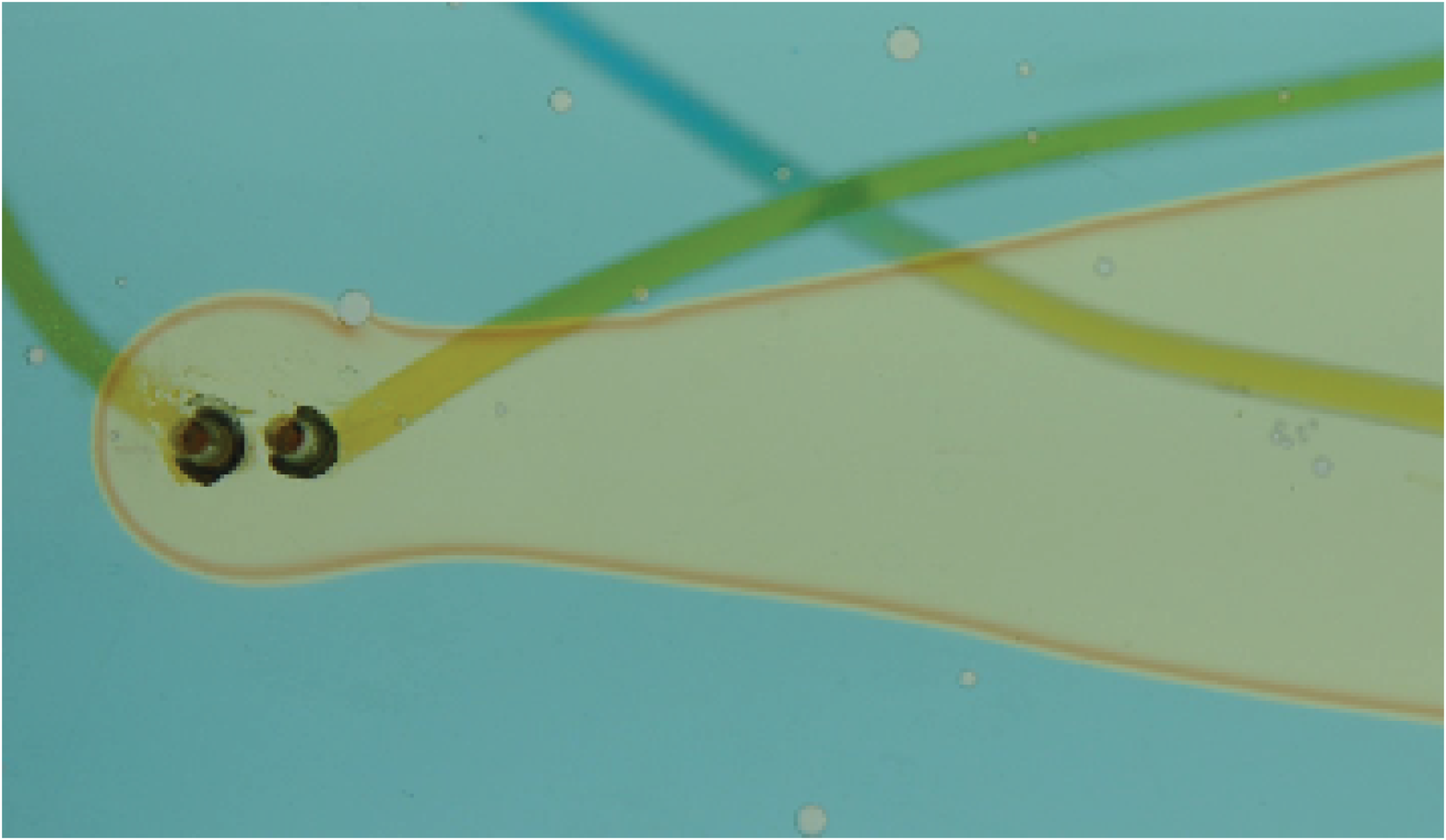}
\includegraphics[width=6cm]{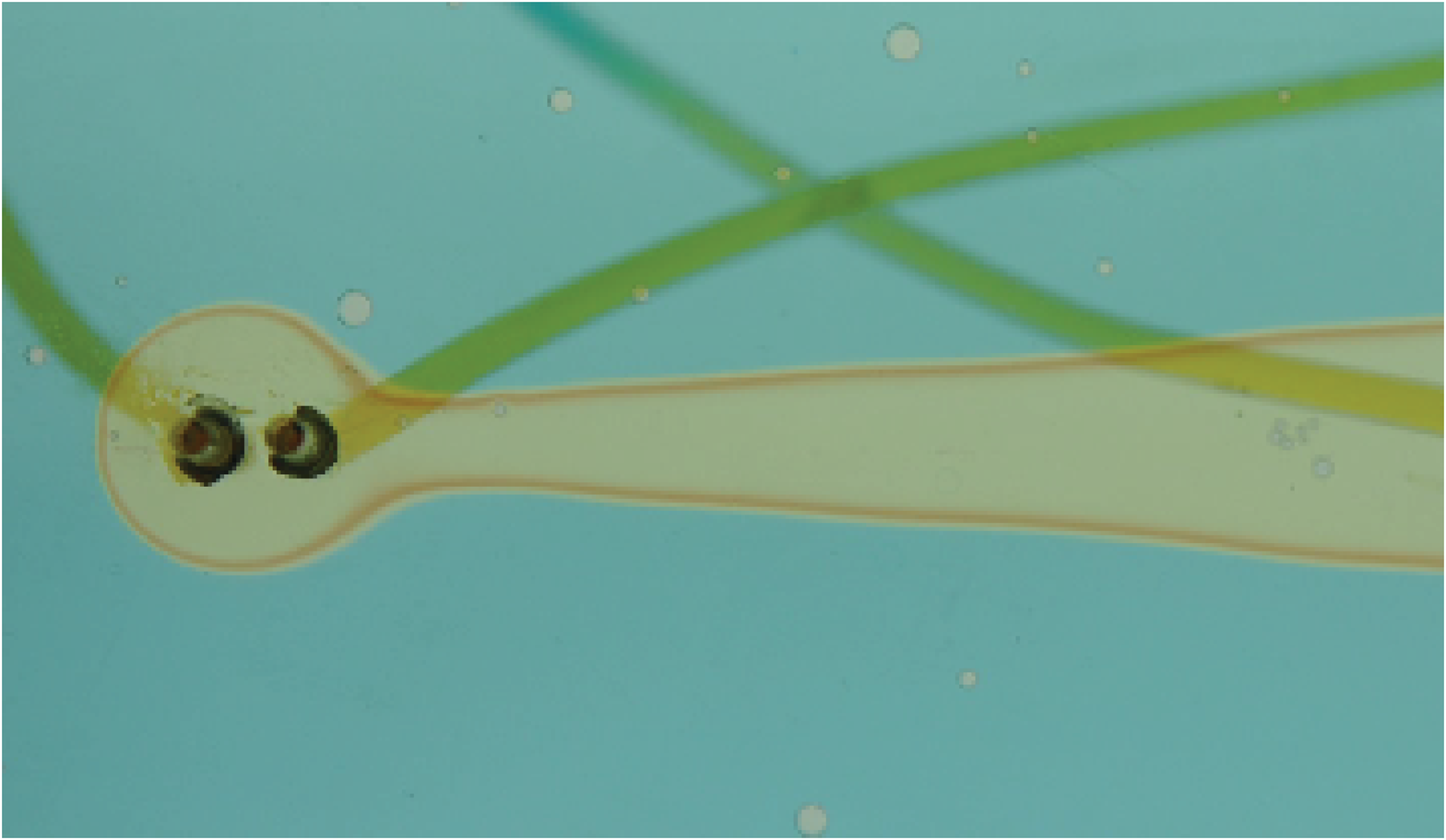}
\includegraphics[width=6cm]{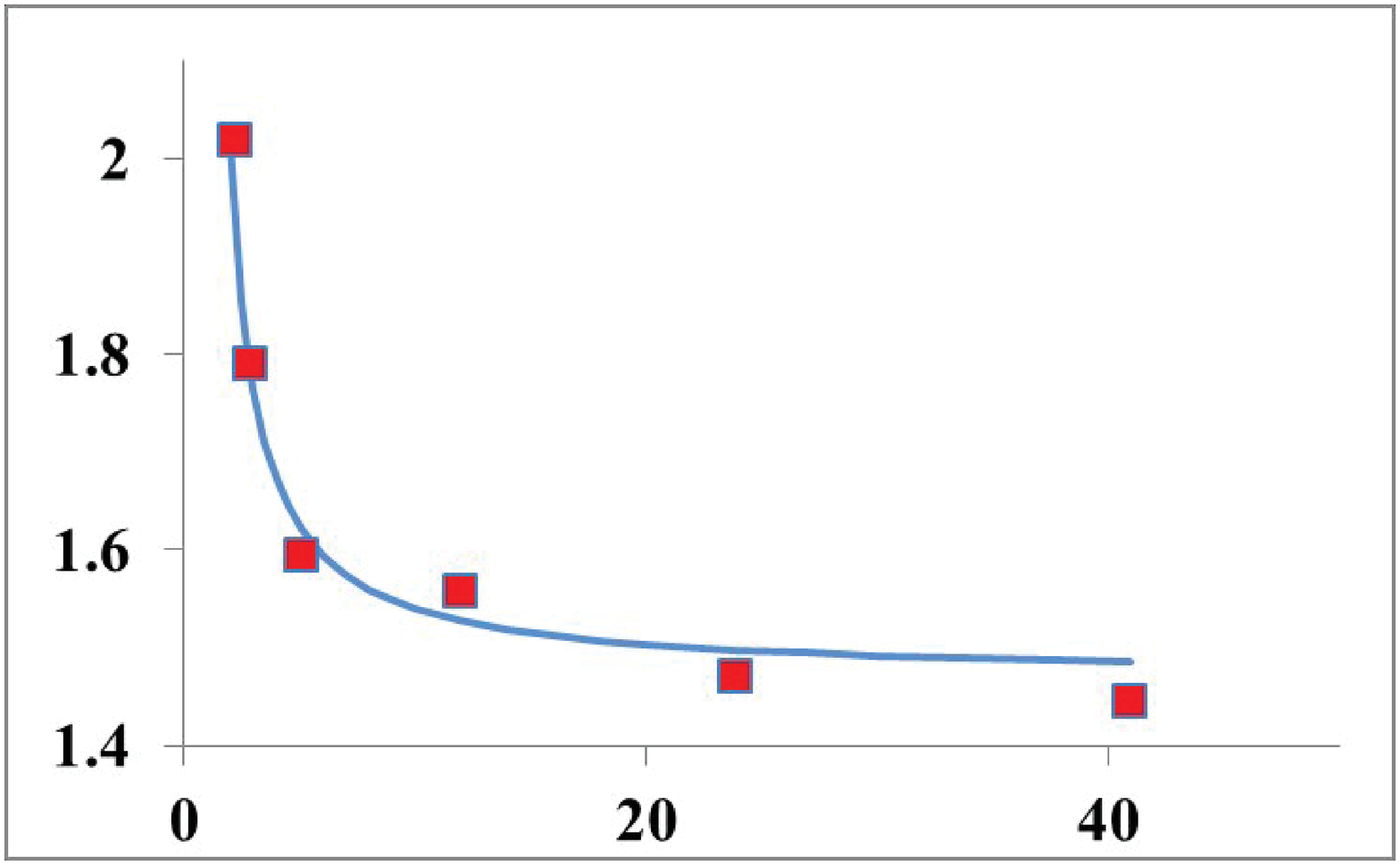}
\caption{\small   Frozen chemical front in a flow around an hydrodynamic dipole. The fresh, blue reactant is injected from left to right as in Fig. \ref{solid}  at a flow at constant velocity $U_0$. The "solid" obstacle is mimicked by an hydrodynamic dipole, injecting burnt yellow product by the leftmost hole (source) and sucking it from the other hole (sink) at the same flow rate ($U_+=-U_-$); the distance between holes is $d=5 \;mm$. In all the pictures, the ratio between $U_0$ and $U_+$ remains the same, hence mimicking the same immaterial obstacle (see text). From top left to bottom right $U_0/V_{\chi}=2,4, 10$. The graph on the right is a plot of the normalized apex, $|h_0|/d$, the closest distance to the dipole on the symmetry axis versus the $u=|U_0/V_{\chi}|$. The continuous line through the data is the selection obtained from the eikonal integration.}
\label{dipole}
\end{center}
\end{figure}
Keeping constant the ratio $U_0/U_+$ provides an immaterial solid disk of constant radius in a uniform flow field $U_0$ \footnote{Indeed, if the hydrodynamic dipole (intensity $p$) is equivalent from the mathematical point of view with a solid disk of radius $R=p/(2 \pi U_0)$, here we have a real source and sink which far field is a dipole. Therefore instead of the dipole flow, we use for the selection calculations the velocity field of a source and a sink distant of $d$  \cite{guyon01} which will be closer to the experimental velocity field especially for the near field.}.

The four pictures on Fig. \ref{dipole} correspond to the Frozen Fronts achieved at different flow rates keeping  $U_0/U_+=2$ constant. The bottom right figure is a plot of the position $|h_0|/d$ of the front on the symmetry axis versus the reduced flow rate $u=|U_0/V_{\chi}|$: the squares are the experimental data the continuous line the theoretical selection obtained using the above paragraph procedure to the fourth order assuming an eikonal equation with the experimental value of $l_{\chi}/d=0.04$: the agreement is quite reasonable validating both the relevance of the eikonal equation in such an experiment and the selection procedure.

\begin{figure}[htbt]
\begin{center}
\includegraphics[width=4cm]{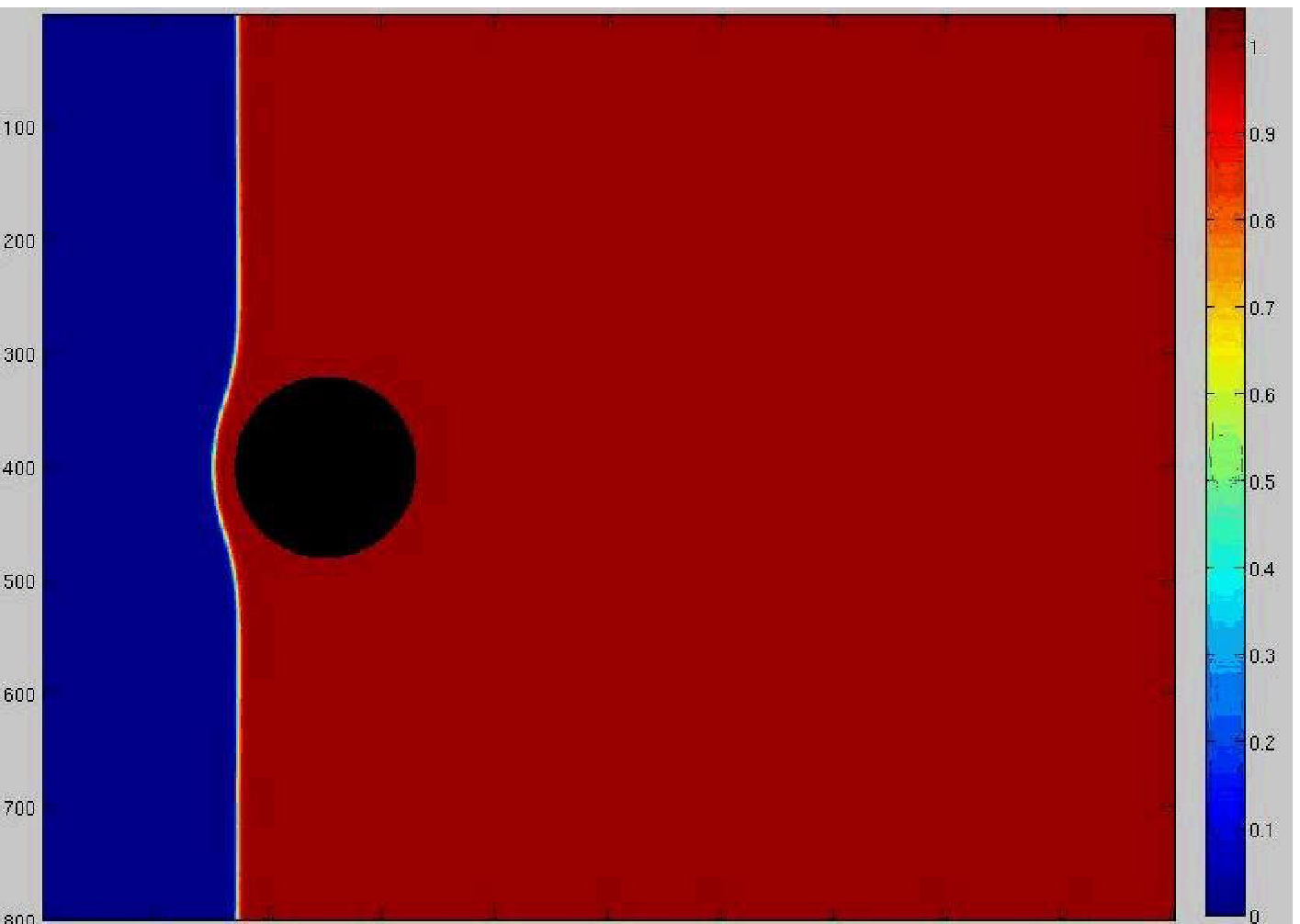}
\includegraphics[width=4cm]{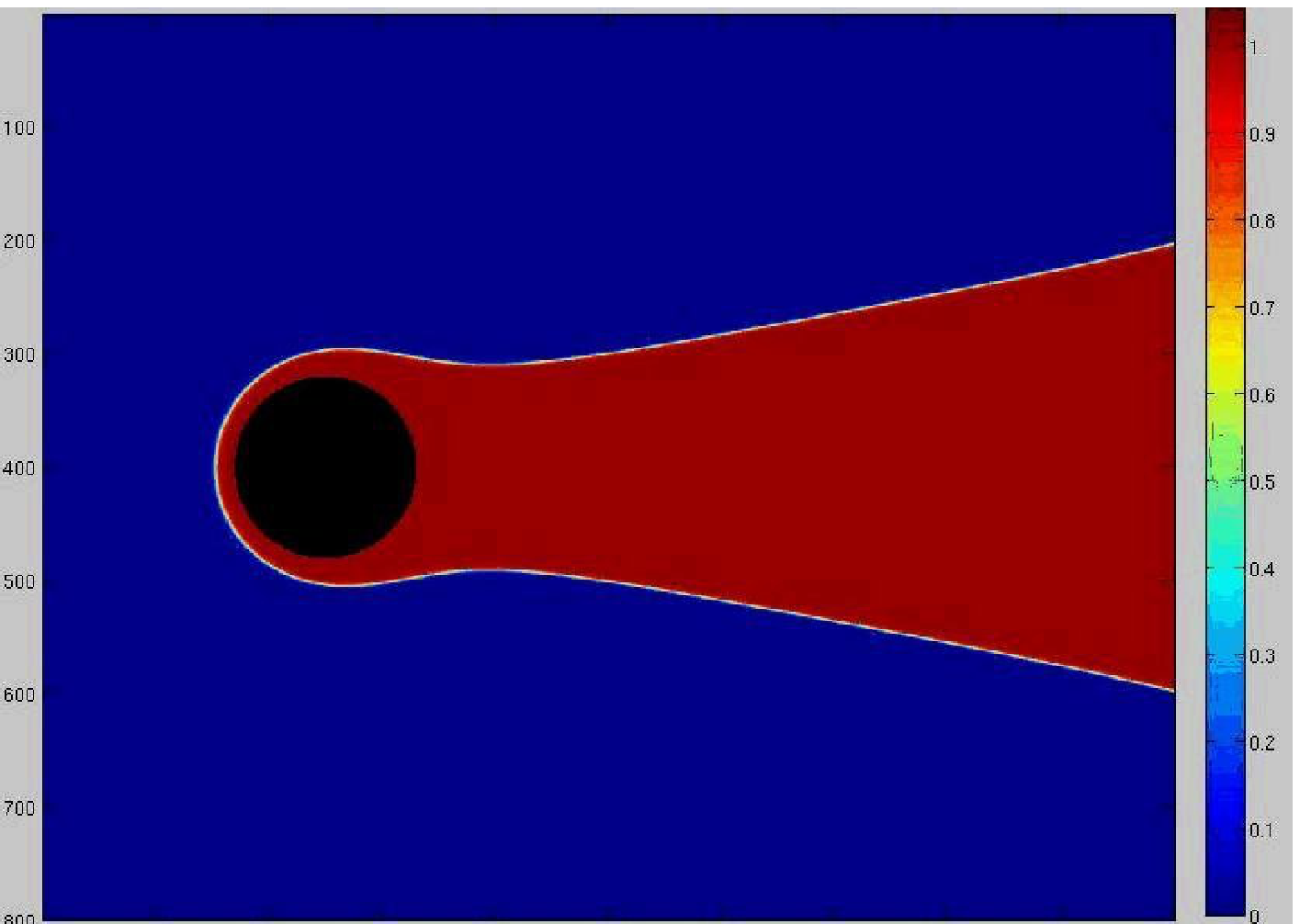}
\includegraphics[width=4cm]{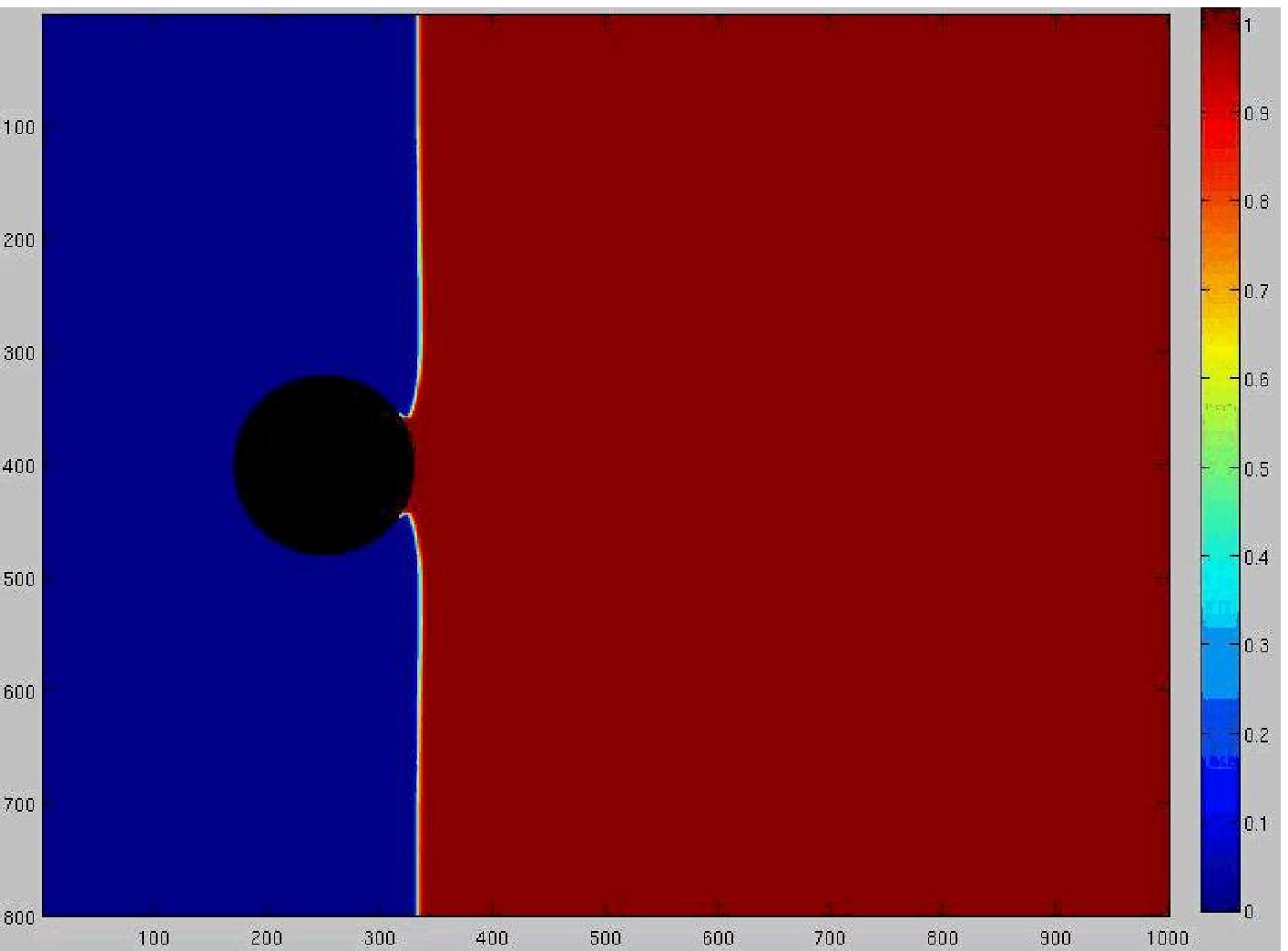}
\includegraphics[width=4cm]{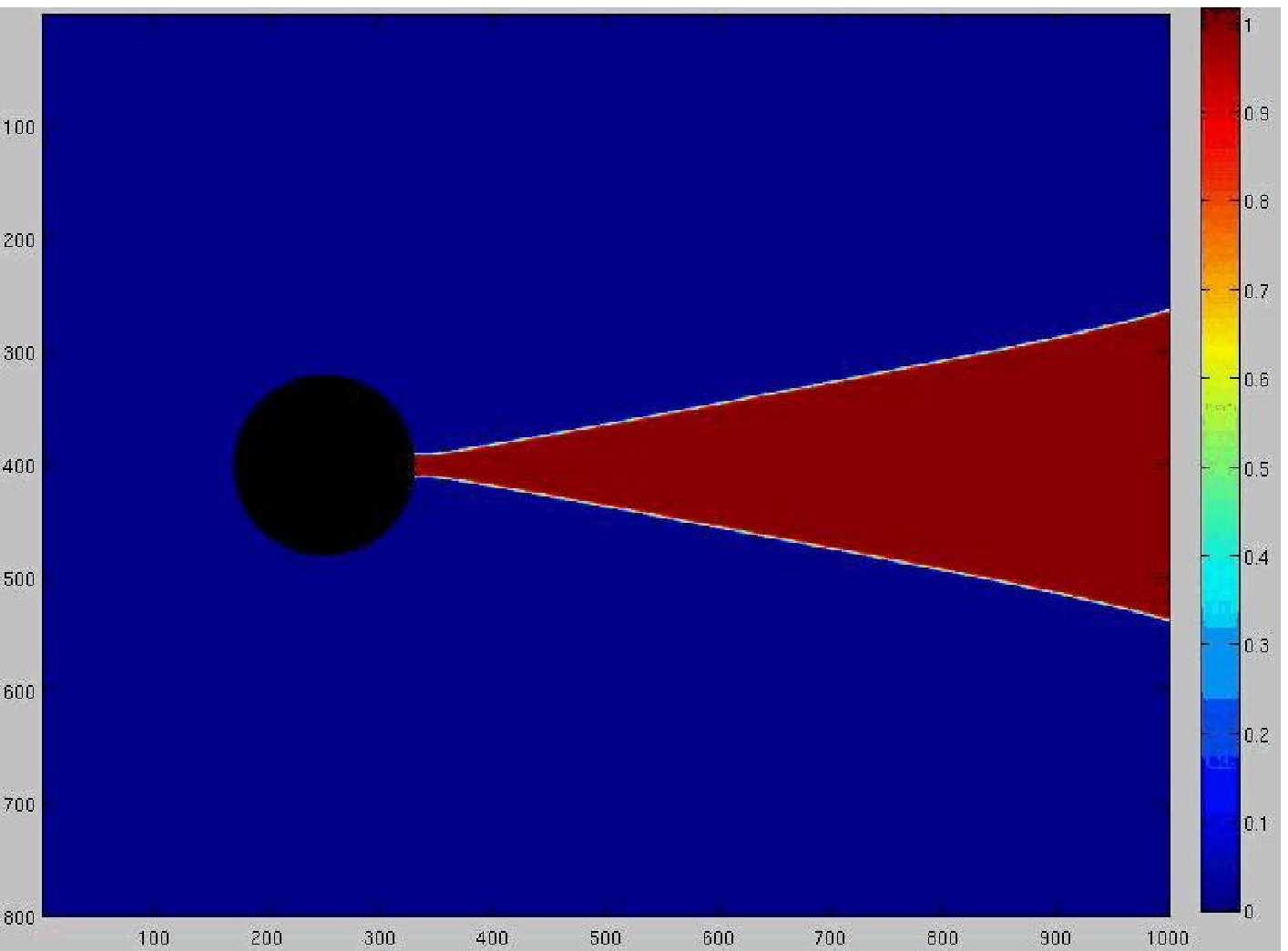}
\includegraphics[width=8cm]{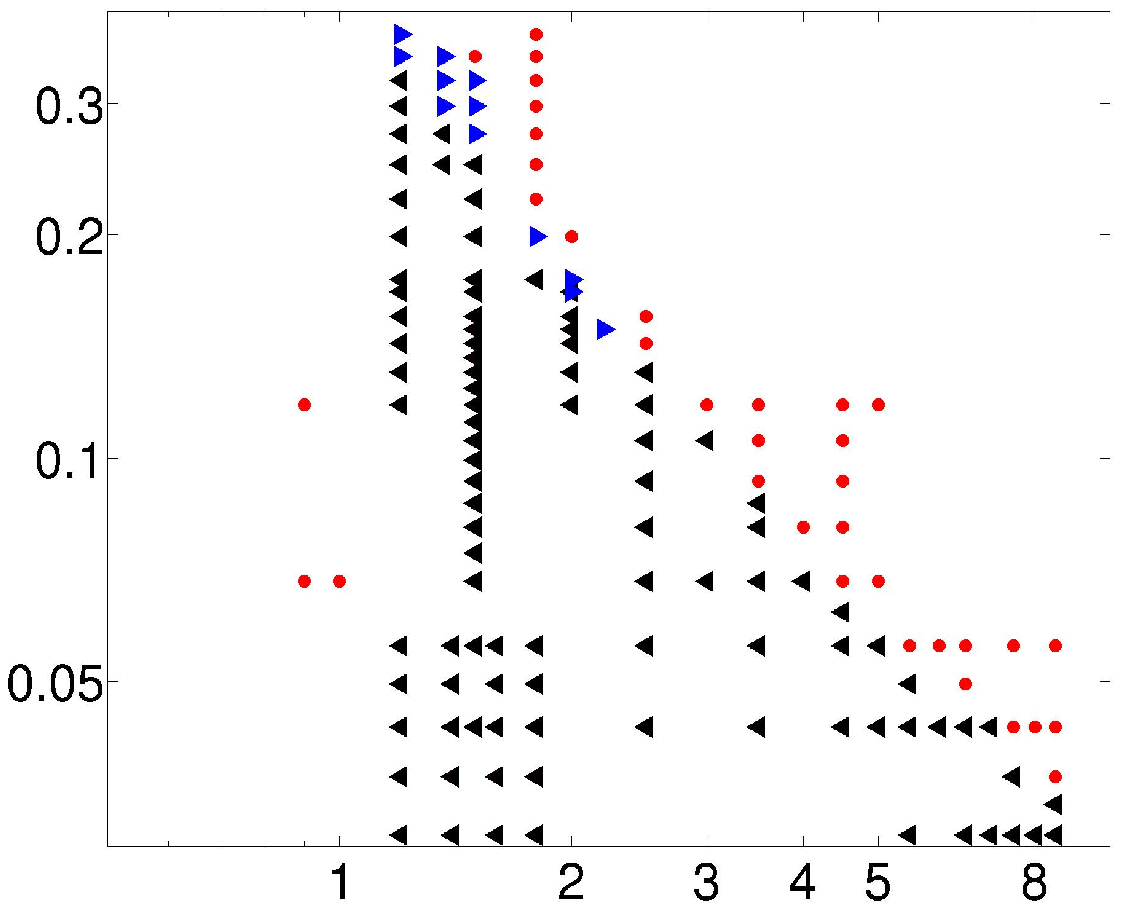}
\caption{\small  Top row : from left to right two sets of the initial condition and the corresponding frozen front. The bottom diagram is a log-log plot of $l_{\chi}/R$ versus $u=|U_0/V_{\chi}|$ in which the red dots ($\bullet$) correspond to unsteady fronts, the black $\blacktriangleleft$ to observation of both upstream and downstream frozen fronts and blue $\triangleright$ to downstream frozen fronts only.}
\label{diagram}
\end{center}
\end{figure}

\section{Phase diagram of different class of Frozen Fronts around a solid disk.}
In the experiments (Fig. \ref{solid}) we have been able, depending on the initial conditions to observe for the same flow rate $u$ two different kinds of Frozen Front: an upstream frozen front avoiding the obstacle and a downstream one in contact with the solid disk. It is worth addressing the issue of the existence of this two kinds of front with the two control parameters $u$ and $l_{\chi}$, that is drawing the phase diagram of these two frozen front types. To cover a wide range of ($u$,$l_{\chi}/R$) values, numerical simulations are more suitable than experiments especially for the control parameter $l_{\chi}/R$. For that purpose, we performed Two Relaxation Times ($TRT$) lattice Boltzmann simulations \cite{ginzburg10}. We use a $2D$  $400 \times 1000$ lattice (Fig. \ref{diagram}) with a solid of size $80$ lattice unit diameter. We first compute the, low Reynolds number, velocity field around this solid disk with periodic boundary conditions at the top and bottom and a constant flux from left to right ($U_0$). We wait until a stationary flow field is achieved. Then we can switch on the reaction with as initial conditions for the reactant/product front, a vertical straight line either upstream (left) of the solid disk (top left in Fig. \ref{diagram}) or in contact with the solid disk on the right (top third picture from the left in Fig. \ref{diagram}). The chosen values of the reaction characteristics ($V_{\chi}$ and $l_{\chi}$) \cite{saha13} fix the two control parameters of the simulation namely $u=|U_0/V_{\chi}|$ and $l_{\chi}/R$  (propagation from right to left in the absence of flow).

\begin{figure}[htbt]
\begin{center}
\includegraphics[width=4cm]{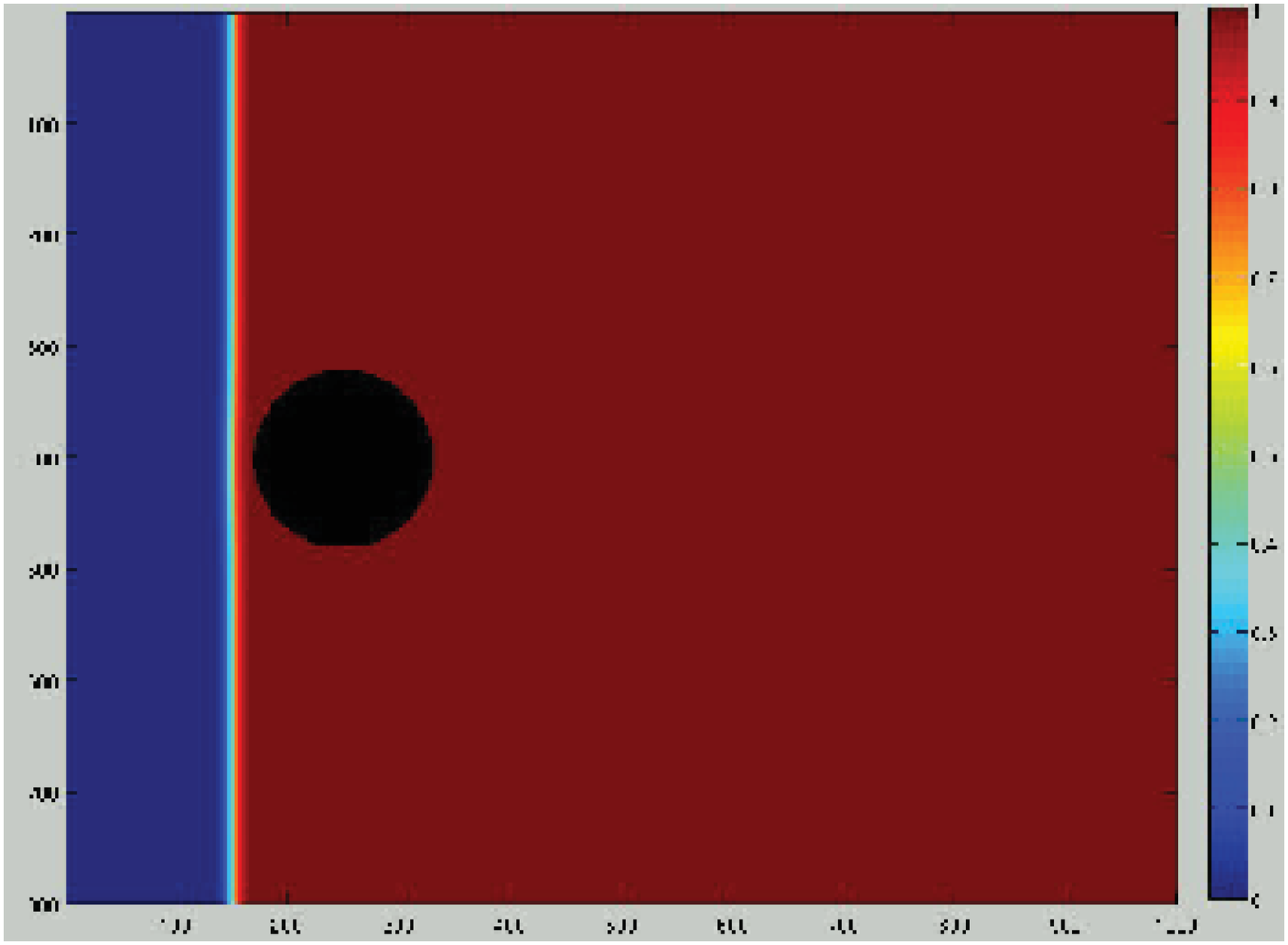}
\includegraphics[width=4cm]{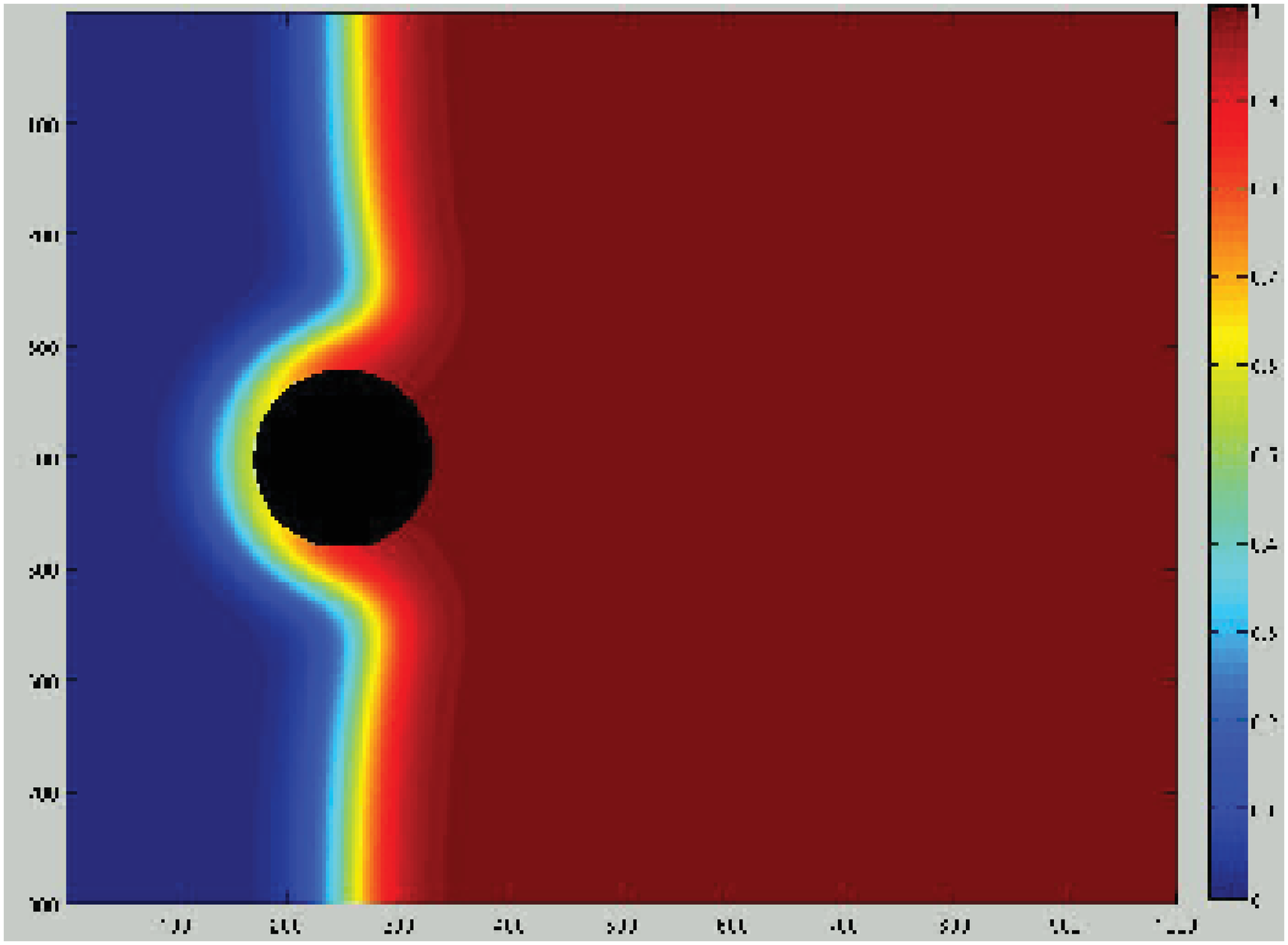}
\includegraphics[width=4cm]{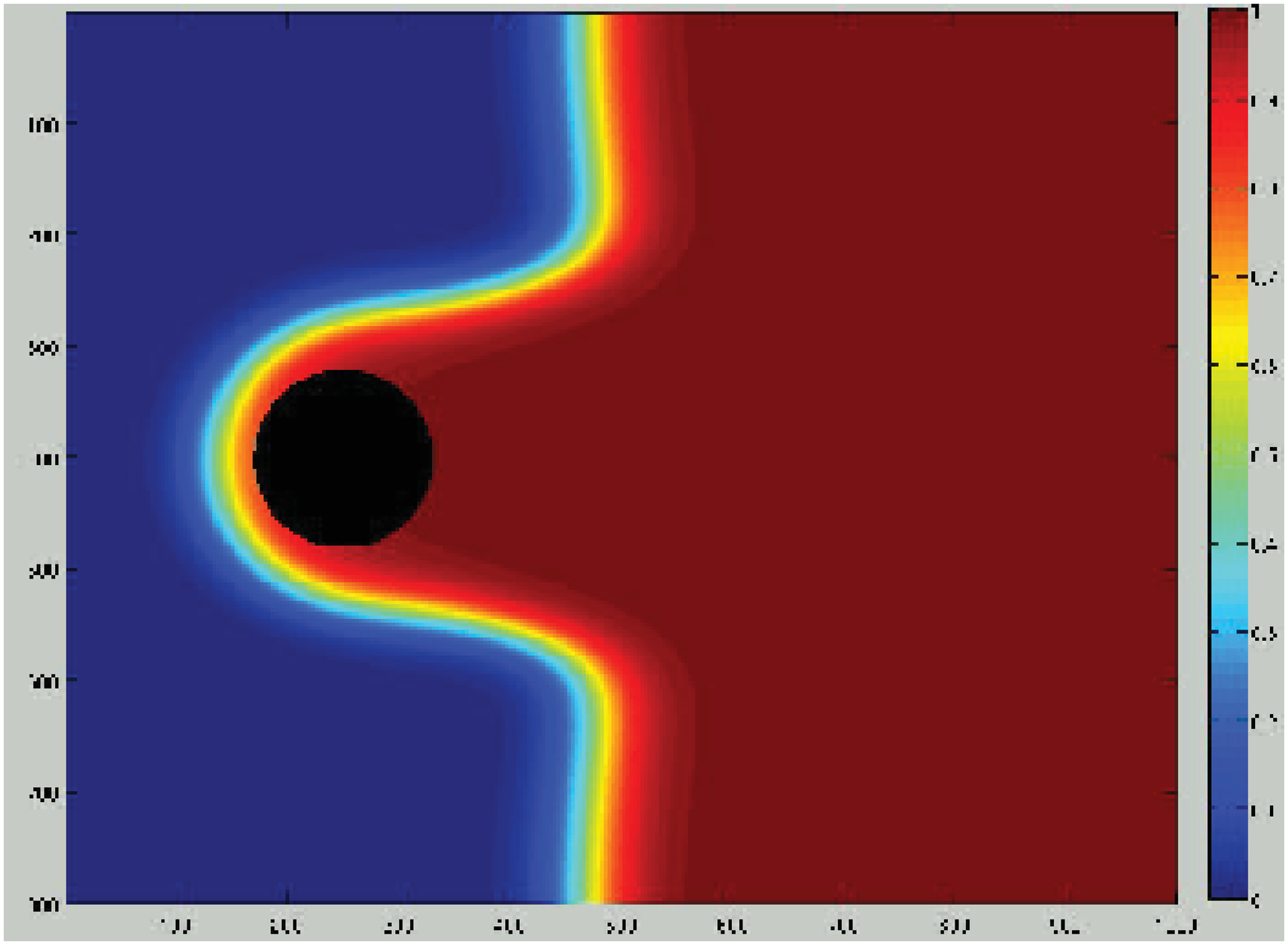}
\includegraphics[width=4cm]{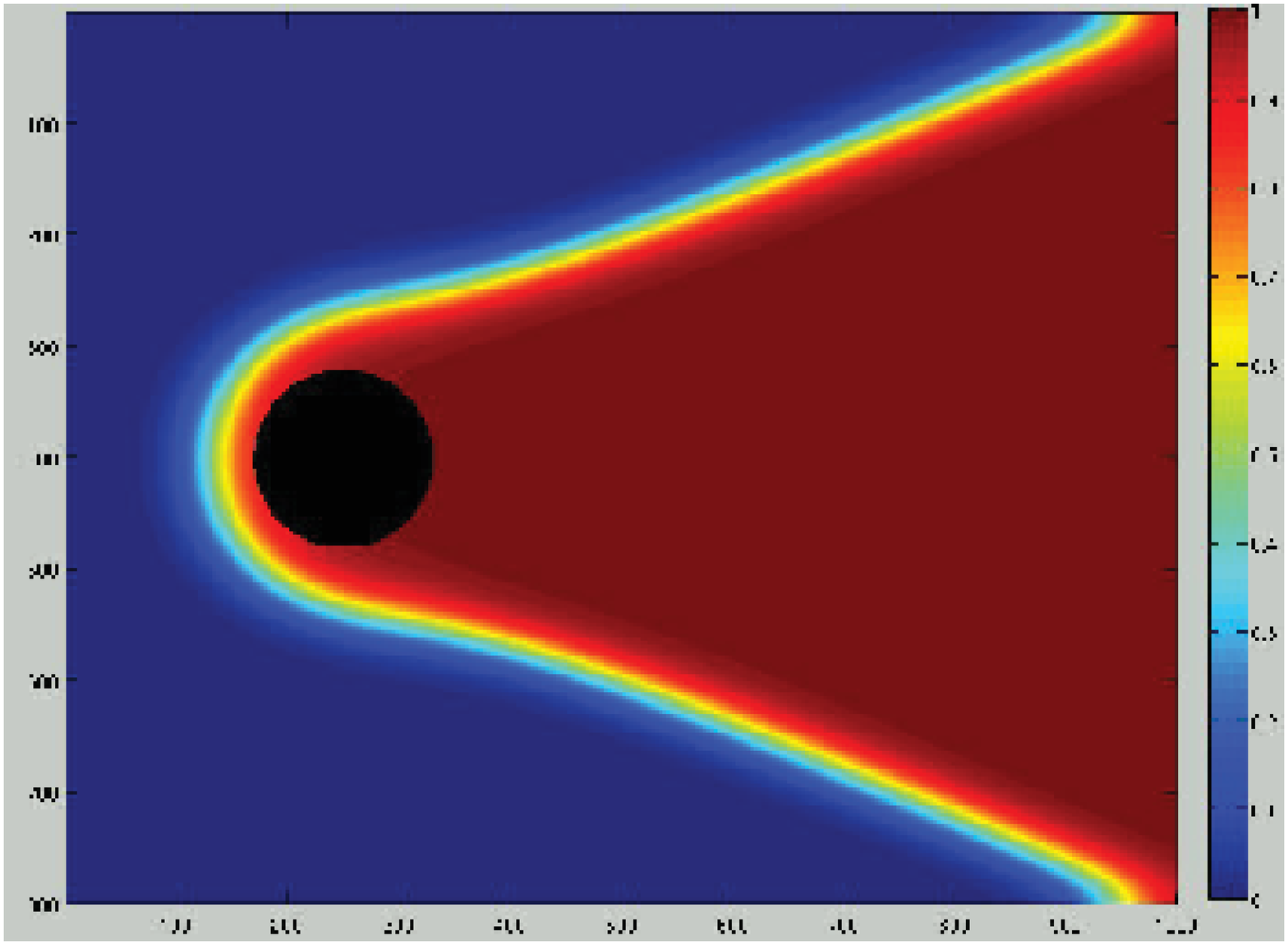}
\includegraphics[width=4cm]{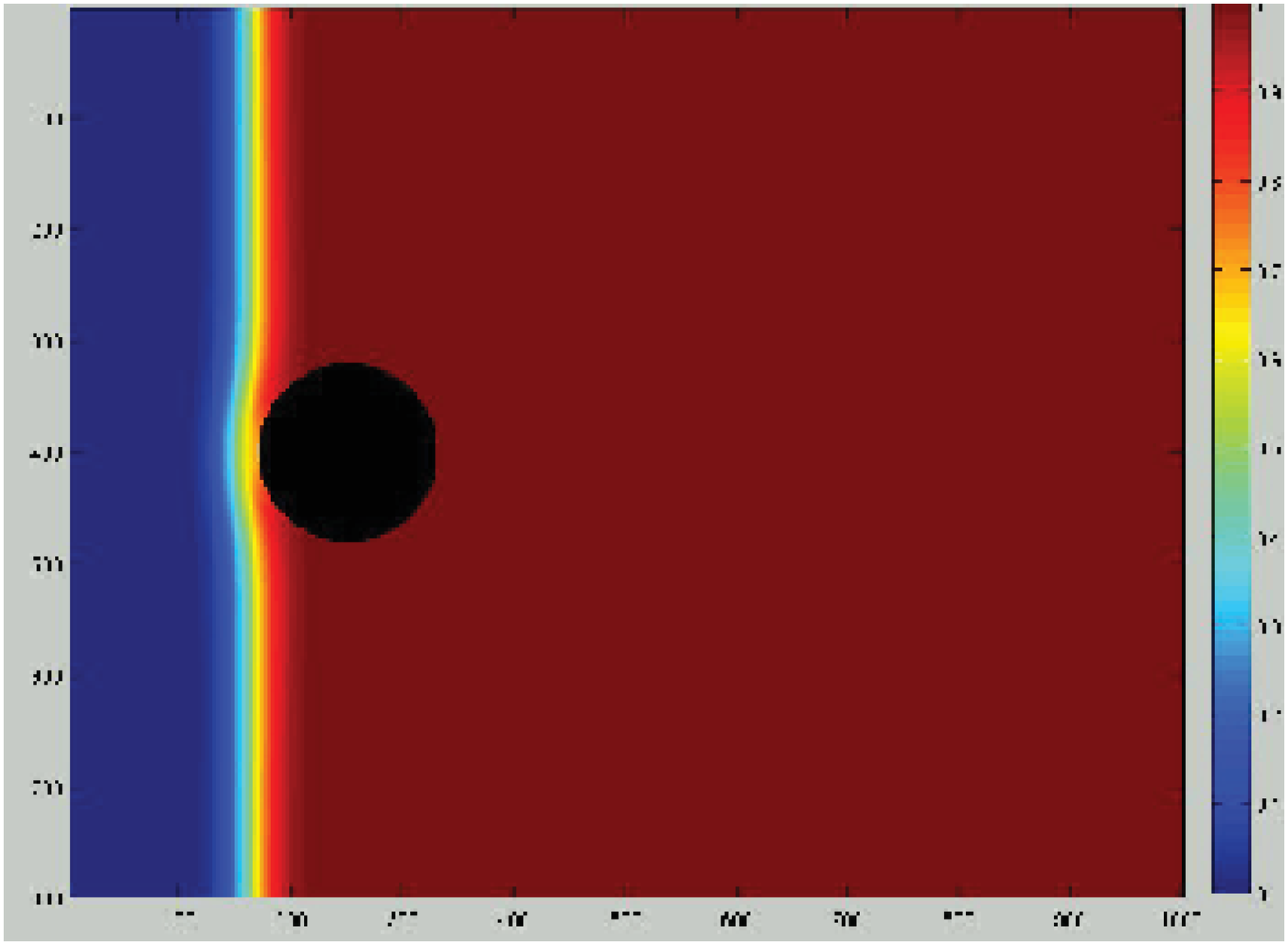}
\includegraphics[width=4cm]{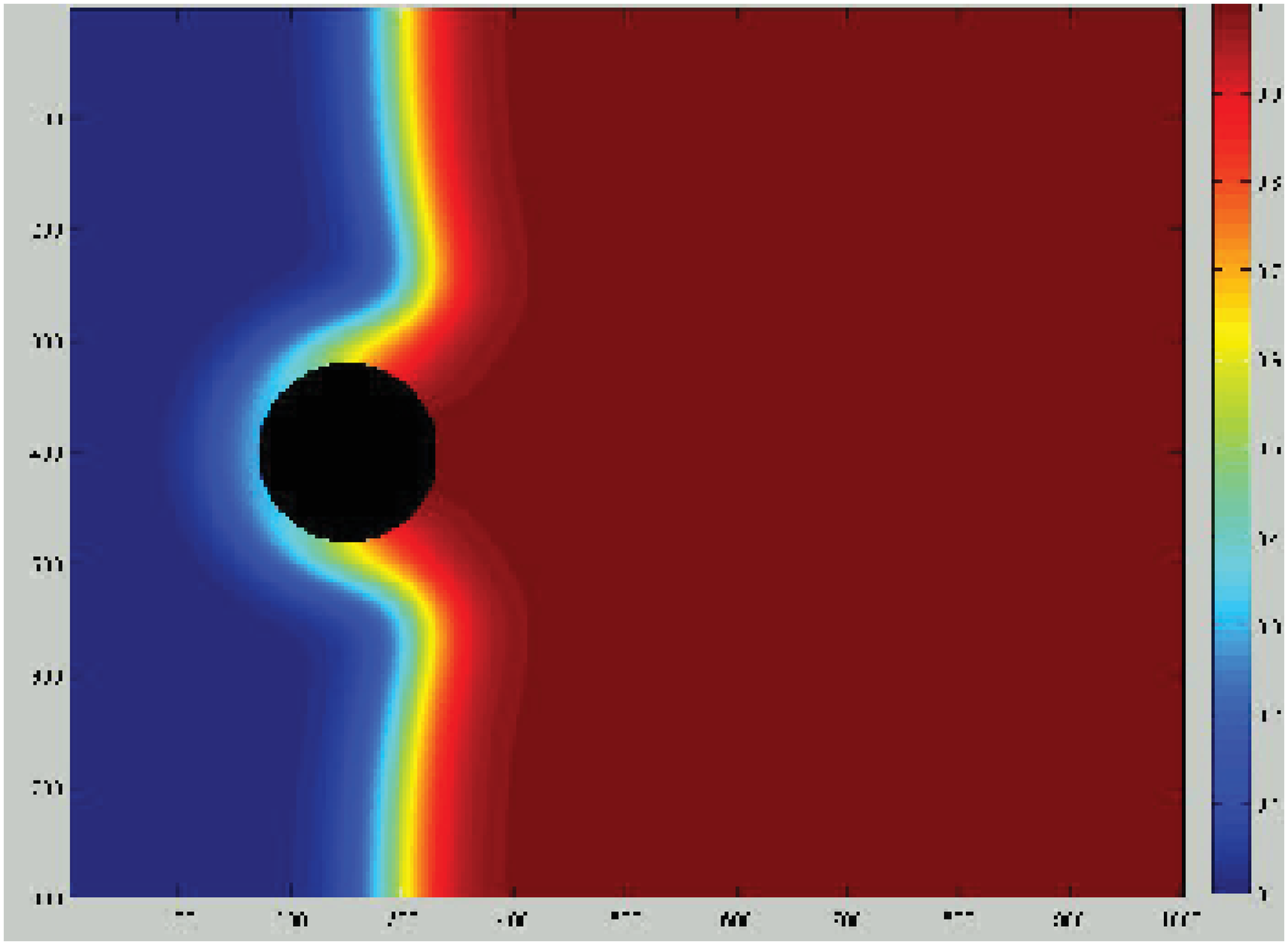}
\includegraphics[width=4cm]{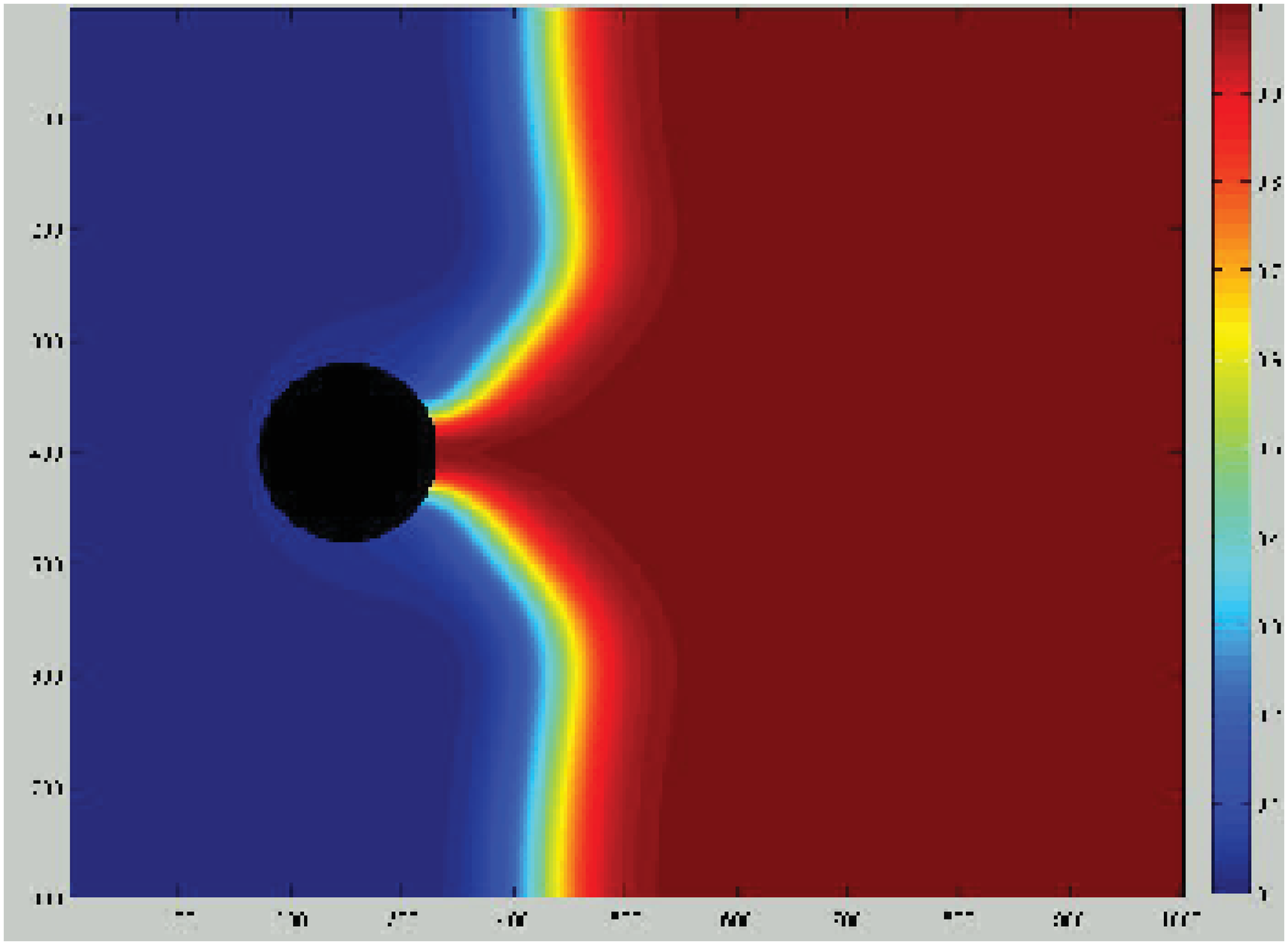}
\includegraphics[width=4cm]{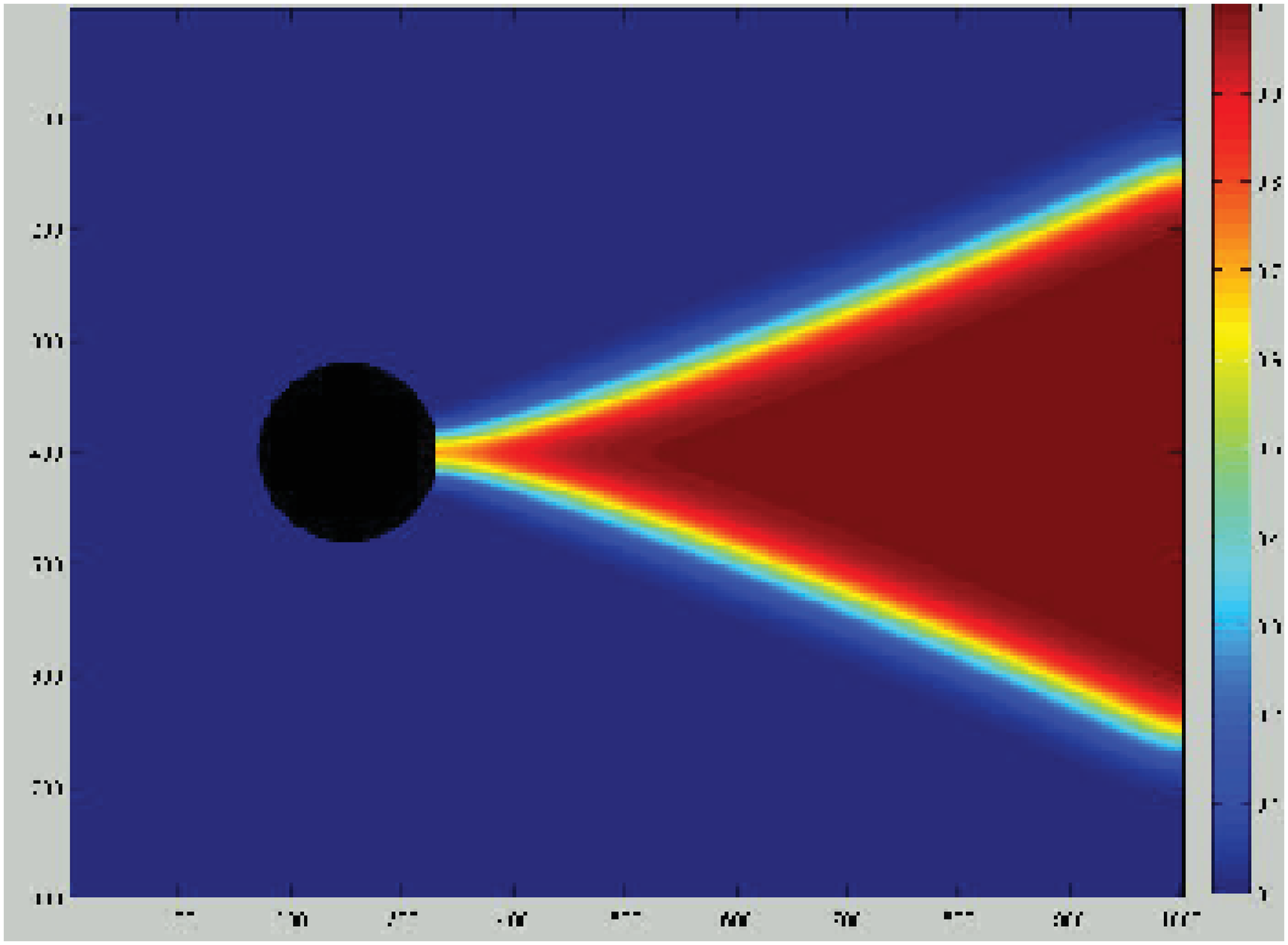}
\includegraphics[width=4cm]{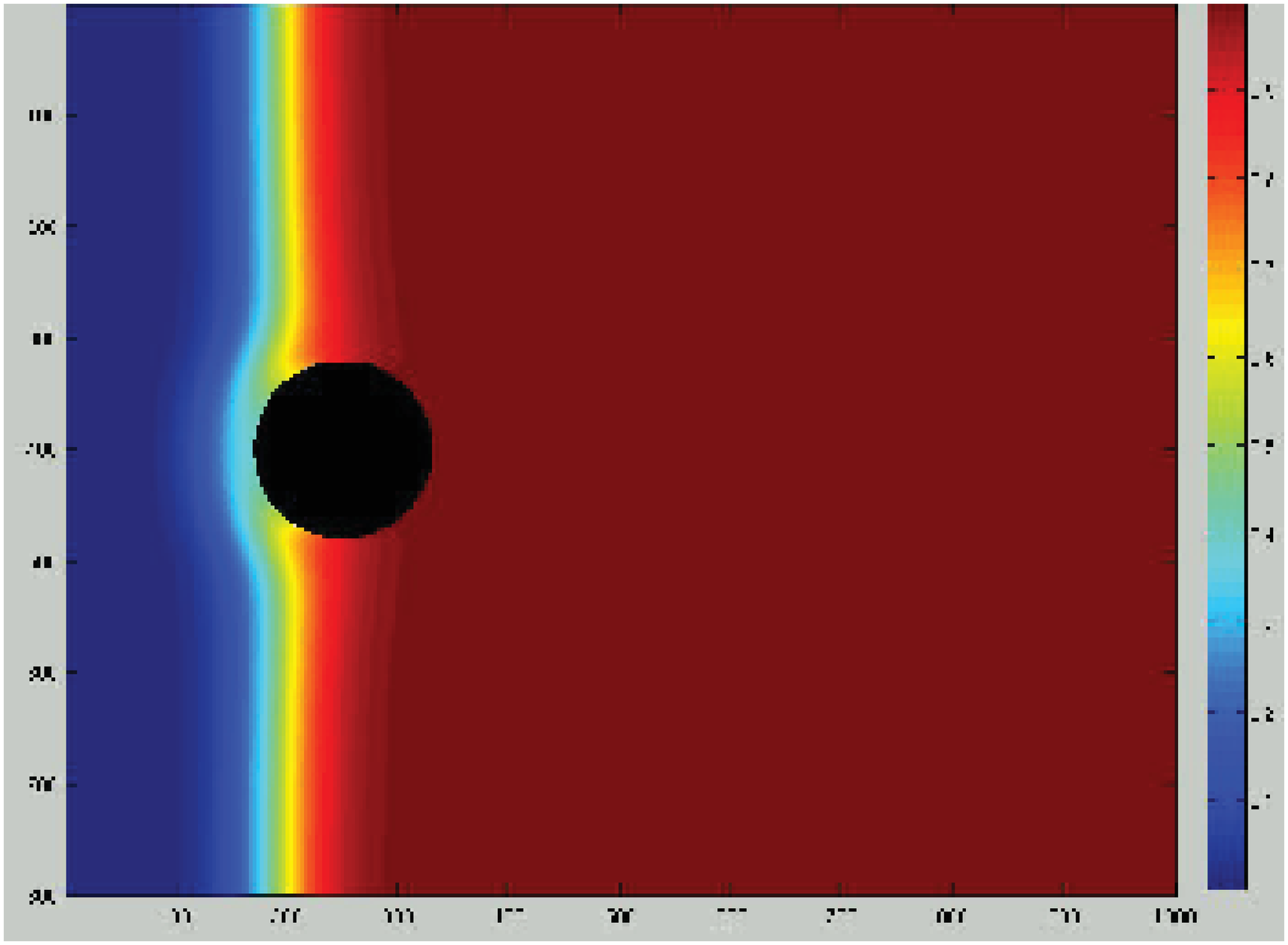}
\includegraphics[width=4cm]{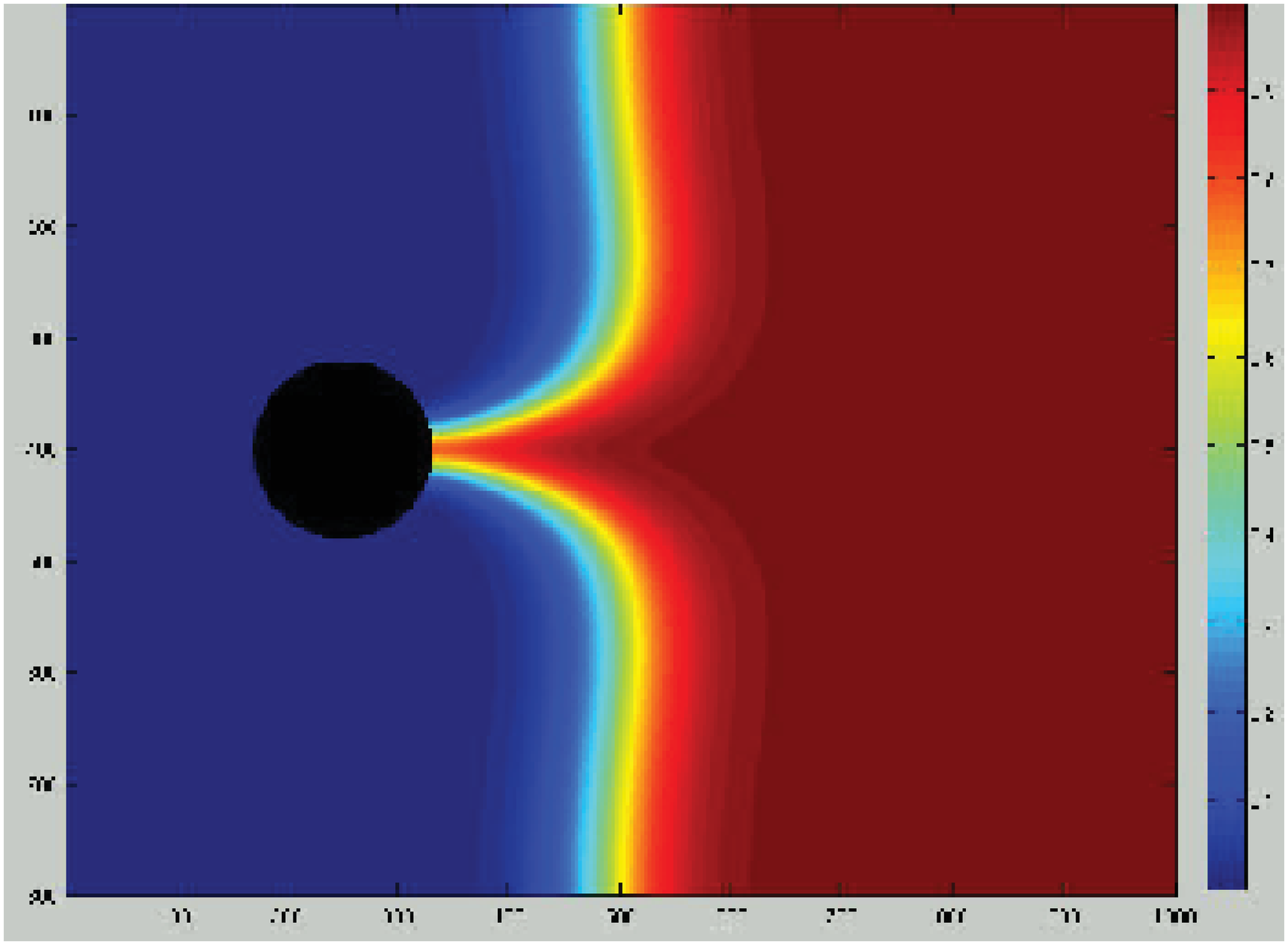}
\includegraphics[width=4cm]{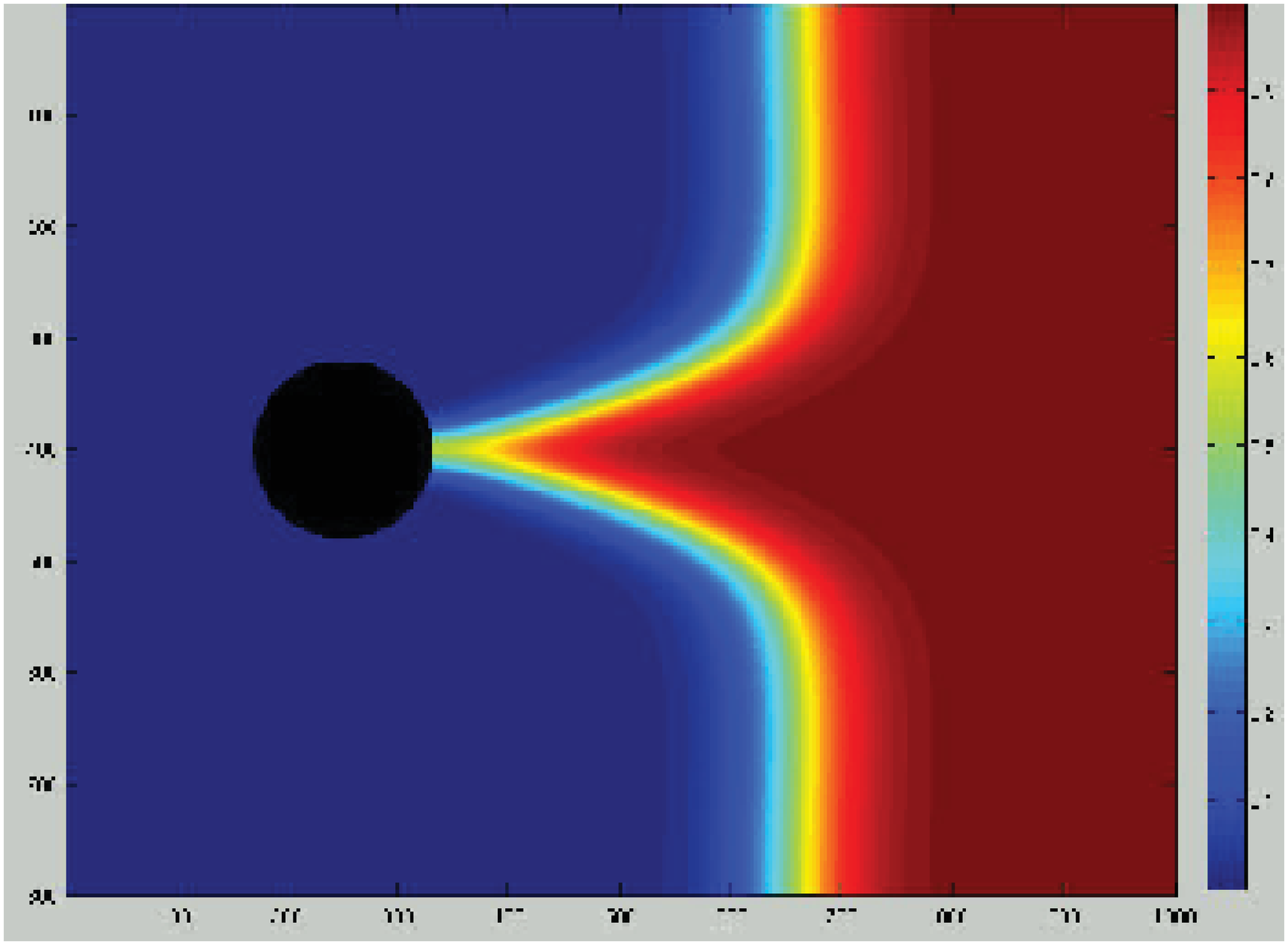}
\includegraphics[width=4cm]{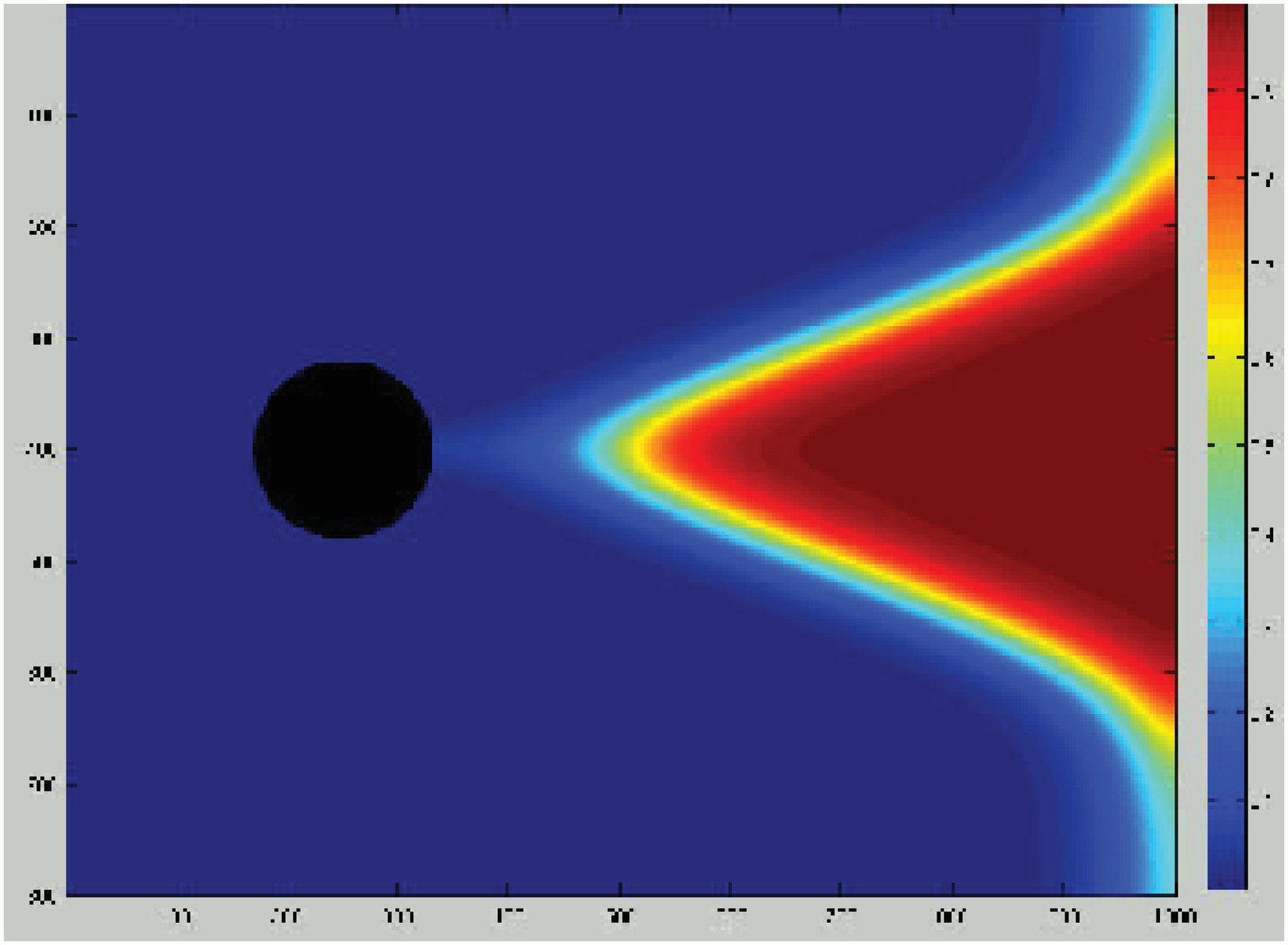}
\caption{\small Successive pictures of a front generated on the left of the disk (upstream conditions) for  $u=1.5$. Top row, $l_{\chi}/R=0.24$ : development of an upstream front. Middle row, $l_{\chi}/R=0.30$ the front fails to froze upstream whereas the front frozes downstream leading  to the downstream frozen front. Lower row, $l_{\chi}/R=0.35$ the front is neither able to froze upstream nor downstream and propagates to the right.}
\label{updown}
\end{center}
\end{figure}

As flow and reaction act, the front develops, deforms and can achieve a stationary, time independent, shape: we observe, depending on the initial conditions, a frozen front either upstream the disk or downstream link on the solid surface as seen in Fig. \ref{diagram}. Therefore our numerical simulations are able to reproduce the experimental observations of the two types of observed frozen fronts. For a constant $l_{\chi}/R$ as the uniform flow velocity $U_0$ is increased, we observe: for $u<1$, there is no stationary front, the front propagates continuously to the left; for $1<u< u_m$ the two types of frozen fronts can be achieved, whereas a further increase of $u$ leads to both fronts detaching and propagating to the right. This is in agreement with the experimental observations, although for a single experimental $l_{\chi}$ value. In the simulations, we can also increase the chemical length at constant flow rate : above a certain $l_{\chi m}$, both types of front become unable to keep stationary. As discussed later on the $u_m$ and $l_{\chi m}$ value can be slightly different for the two types of fronts. There is a hierarchy between the stability of the upstream and downstream fronts : for all the many values of ($u$,$l_{\chi}/R$), when a upstream front is observed there is also a corresponding downstream one (black $\triangleleft$) in Fig. \ref{diagram}. In a small window of ($u$,$l_{\chi}/R$) values we do observe only the downstream frozen fronts (blue $\triangleright$). This larger stability of the frozen front is demonstrated on the middle row of Fig. \ref{updown} where the upstream front barely try to froze upstream around the solid disk, but failed whereas it succeeds downstream. The diagram on Fig. \ref{diagram} is a log-log plot of $l_{\chi}/R$ versus $u=|U_0/V_{\chi}|$ in which the red dots correspond to unsteady fronts. This diagram clearly shows that increasing either $u=|U_0/V_{\chi}|$ or $l_{\chi}/R$, keeping the other constant, leads to the lack of frozen fronts. This can be easily at least qualitatively understood: increasing the flow velocity results in a frozen front closer to the disk for the upstream front and closer downstream branches for the downstream one. As a result the velocity increase promotes the front touching the disk from the left or the two downstream branches to merge. We observe the same effect by increasing $l_{\chi}$. To be more quantitative, on the log-log plot of Fig. \ref{diagram} the boundary between frozen fronts and unsteady fronts is almost straight line of slope $\sim-1$, that is $u \sim 1/ (l_{\chi}/R)$. The argument to account for this slope is as follows. The front extension on the symmetry axis is roughly $|h_0|\pm l_{\chi}$, the contact with the disk and hence the lack of frozen front corresponds to  $|h_0|+l_{\chi}\sim R$. As the flow velocity on the axis,  $v_x(h_0,0)\sim u(1-(R/h_0)^{2})$ is balanced by the chemical velocity including the curvature ($\sim 1/R$) effect ($1-l_{\chi}/R$) leads to

\begin{equation} \label{slope}
u(2+x)x=(1-x)(1+x)^{2}  \;\; {\text with} \;\; x=l_{\chi}/R
\end{equation}
which, for not too large $x$, is $u\simeq \frac{1}{2 x}=\frac{R}{2 l_{\chi}}$, in agreement with the slope on the diagram.

\begin{figure}[htbt]
\begin{center}
\includegraphics[width=10cm]{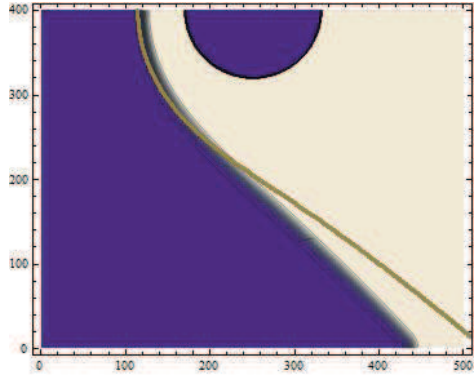}
\caption{\small Superposition on the same graph of the iso-concentration map from the simulation and of the eikonal front (solid curve) for $u=1.5$ and $l_{\chi}/R=0.037$.}
\label{EikSimul}
\end{center}
\end{figure}

\section{Simulation with a finite $l_{\chi}$ versus eikonal}
From the previous section, we have accurate numerical simulations of the frozen fronts, $C(x,y)$ for different chemical length $l_{\chi}$ (and flow rates). Therefore it deserves to compare these simulations with the eikonal limit obtained in the corresponding section.
Fig. \ref{EikSimul} is the superposition of the iso-concentration map from the simulation and of the eikonal curve for $U_0=-1.5V_{\chi}$ and $l_{\chi}/R=0.037$. Even though for such a small $l_{\chi}$ we are already in the plateau \ref{selection} of the full eikonal limit, there not a matching between the two. Of course we cannot expect that the single concentration jump eikonal curve match with the $C=0.5$ iso-concentration of the simulation but the difference is surprising, especially far away from the solid disk where the velocity field is uniform  $\overrightarrow{U}=\overrightarrow{U_0}$. Let us try to understand this difference in this uniform flow field. The eikonal is a straight line with an angle $\theta_{0}$ with the x axis, along $\overrightarrow{U_0}$. The eikonal equation \ref{eikonal}, $\overrightarrow{U_0} \overrightarrow{n}+V_{\chi}=0$ reads as $U_0 cos(\theta_0)=V_{\chi}$. On Fig. \ref{EikSimul}, we measure $\theta_0=48.5^{\circ}$ in agreement with the eikonal expectation: $cos(\theta_0)=0.66\simeq 1/1.5 $.
\\ A fine analysis of the simulations shows that the iso-concentration are quite perfectly straight lines parallel to one another with an angle $\theta > \theta_0$ with the x axis. In the direction $z$ perpendicular to these iso-concentrations, the concentration is fitted almost perfectly by an equation similar to the chemical front in the absence of flow (Eq. \ref{eq:vl}): $C(z)=\left(1+\exp-(\frac{z}{w})\right)^{-1}$, but with a width $w=2.49$ smaller than $l_{\chi}=3$. As the iso-concentration are parallel straight lines the steady convection diffusion reaction Eq. \ref{CRD} reduces to

\begin{equation} \label{isoconc}
U_z \frac{d C}{d z}=\frac{1}{\tau} C^{2}(z)(1-C(z))=\frac{2 V_{\chi}}{l_{\chi}}C^{2}(1-C)
\end{equation}
where $U_z=U_0 cos(\theta)$ is the projection of the uniform velocity on the normal to the iso-concentrations. Integration of this equation along $z$ from $-\infty$ to $\infty$  corresponds to the balance between convective flux and reaction, leading to
\begin{equation} \label{sl}
U_z=U_0 cos(\theta)=V_{\chi}\frac{w}{l_{\chi}}
\end{equation}
As $w$ is smaller than $l_{\chi}$, $\theta > \theta_0$ is in agreement with what is observed in Fig. \ref{EikSimul}:  $\theta=56.5^{\circ}$ leading to $cos(\theta)=0.55\simeq 2.49/(3\cdot 1.5)$ is in agreement with Eq. \ref{sl}. These observations deserve comments. Even if the  chemical length is very small compared to the disk size ($l_{\chi}R\simeq0.037$), we are not in the full eikonal regime $l_{\chi}\rightarrow 0$ as already observed \cite{leconte04}. The extension of the front, $\sim l_{\chi}$ still matters; in this uniform flow region, the iso-concentrations adapt their spreading as well as their orientation to fulfill the convection reaction balance, the eikonal curve having only the orientation freedom.

\section{Free boundary conditions at the disk surface: Experiments with an air bubble.}
The experiments with the dipole design has proved the relevance of the eikonal to account for the shape of Frozen Fronts around an immaterial obstacle. The problem would not be so easy for a front originating from the solid surface. In the eikonal limit the front cannot be static: at the solid surface the flow velocity vanishes $\overrightarrow{U}=\overrightarrow{0}$ and unless the curvature at the surface is of the order of $l_{\chi}$, the only possibility left from Eq. \ref{eikonalpropa} is $\overrightarrow{V_F}=\overrightarrow{V_{\chi}}$ as already discovered \cite{edwards02}.

\begin{figure}[htbt]
\begin{center}
\includegraphics[width=4cm]{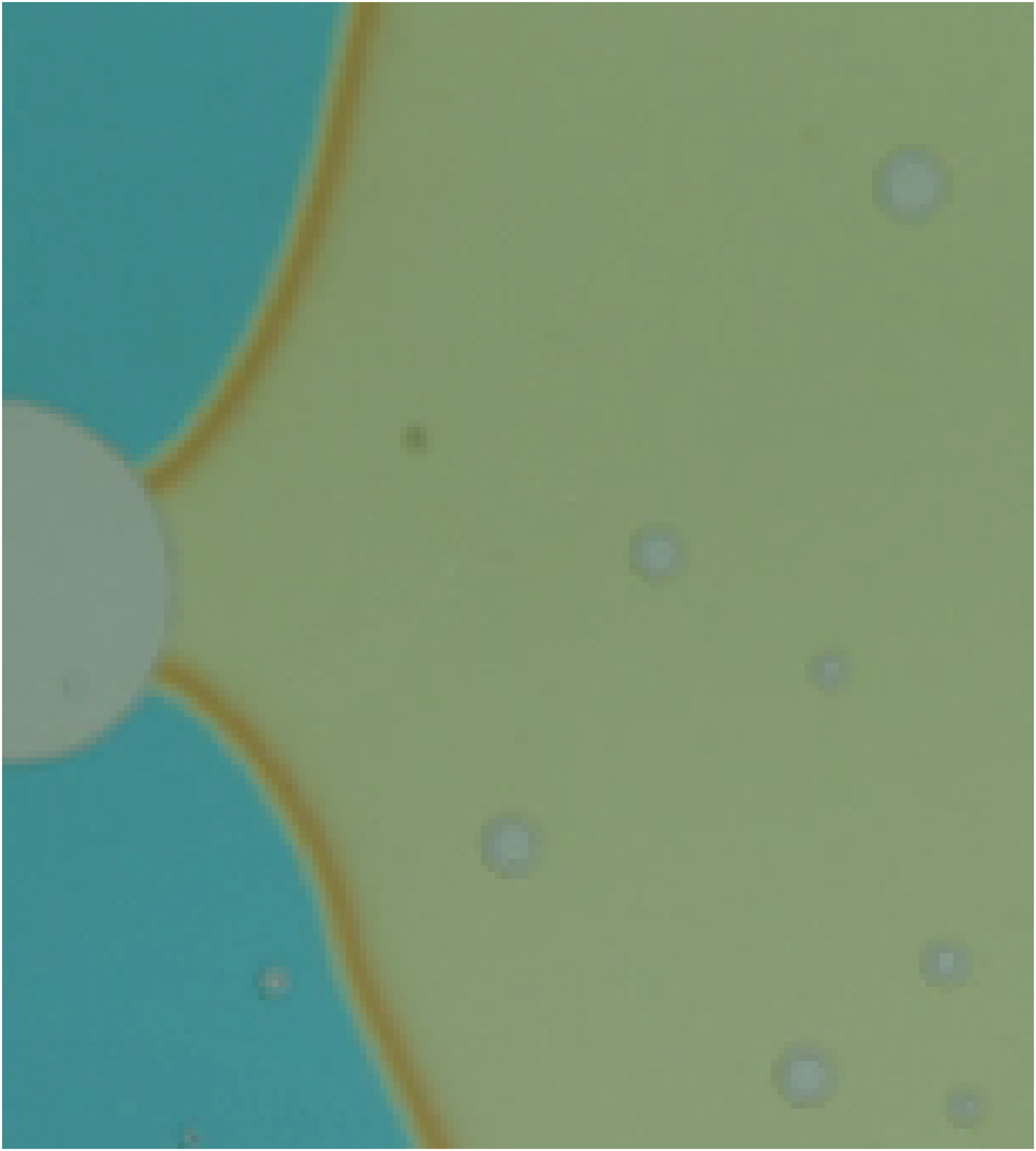}
\includegraphics[width=4cm]{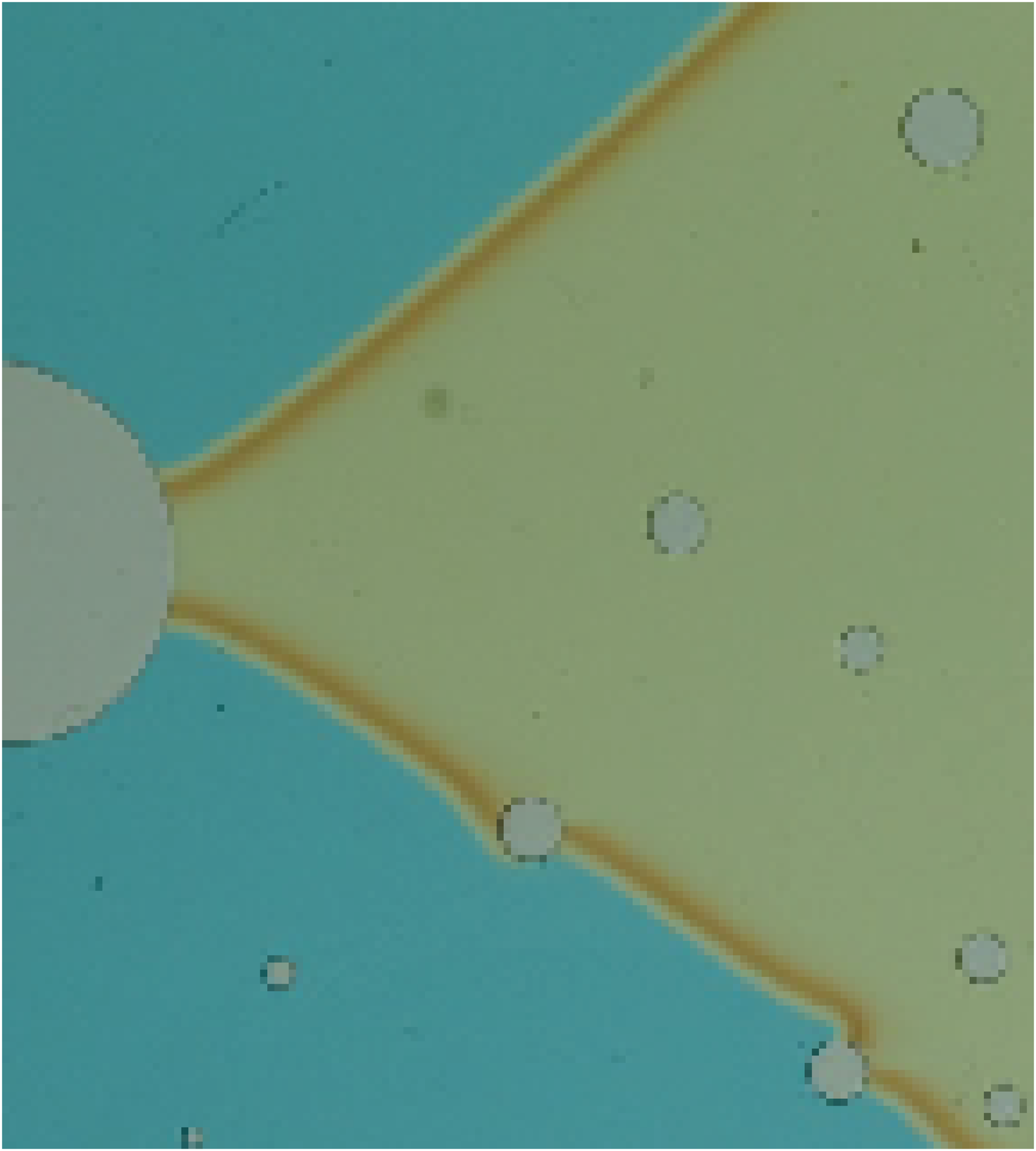}
\includegraphics[width=4cm]{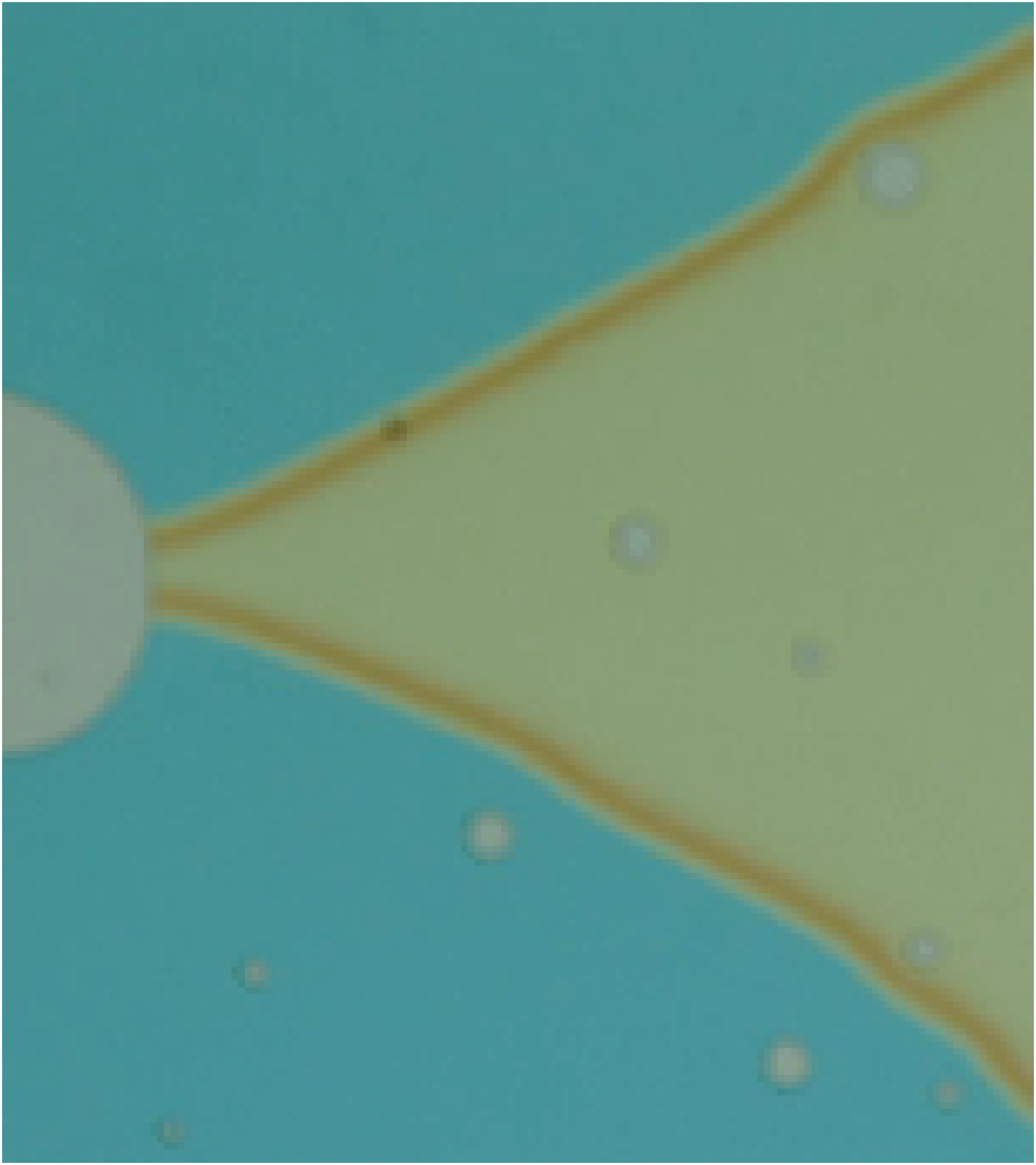}
\caption{\small  Frozen chemical front in a flow around an air bubble of diameter $10 \; mm$ in a Hele-Shaw cell at different flow rates. Compared to Fig \ref{solid}, where the disk is solid, the front is perpendicular to the bubble surface. From left to right $u=1.2$, $u=2$ and $u=4$.}
\label{bubble}
\end{center}
\end{figure}

To get rid of this lack of velocity at the disk surface, we perform experiments with an air disk, an air bubble quenched between the two plates of the Hele-Shaw cell: the boundary conditions between these two fluids (air and chemical solution) are the previous one of zero normal velocity ($v_r(R)=0$) but  the free boundary allows a non zero tangential velocity $ v_{\theta}$ at the bubble surface \cite{lamb32}. Moreover, at the bubble surface, there is no flux of matter ($\overrightarrow{\nabla}C.\overrightarrow{n}=0$) leading to a front perpendicular to the surface. Therefore the eikonal equation at the bubble surface reduces to : $v_{\theta}|_{R}=V_{chi}$. Fig. \ref{bubble} show pictures of the front on a bubble for different flow rates. Compared to the solid disk Fig. \ref{solid}, at the bubble surface, the front is perpendicular to the bubble as expected from the eikonal equation. To be more quantitative on the relevance of the eikonal equation, we have measure the slopes (or the angle with the symmetry axis) of the straight part of the front, at the bubble surface (angle $\theta$) and far away from the bubble ($\alpha$) where the flow is uniform ($U_0$). These measurements are in reasonable agreement with the eikonal expectations : in the far field $U_0 sin(\alpha)=V_{\chi}$ and at the bubble surface where the velocity $v_{\theta}$ is given from the potential flow approximation \cite{lamb32}, $2 U_0 sin(\theta)=V_{\chi}$.

\section*{Conclusions}

We perform experiments, analytical computations and numerical simulations with the autocatalytic Iodate Arsenious Acid reaction ($IAA$) over a wide range of flow velocities around a solid disk. For the same set of control parameters, we observe two types of frozen fronts: an upstream $FF$ which avoid the solid disk and a downstream $FF$ with two symmetric branches emerging from the solid disk surface. We delineate the range over which we do observe these Frozen Fronts. We also revisit the so-called eikonal, thin front limit to describe the observed fronts and to select the frozen front shapes.\\

{\it Acknowledgements.-} \ni
It is a pleasure to acknowledge Agence Nationale de la Recherche for financial support of the project LaboCothep ANR-12-MONU-0011.



\begin{thebibliography}{27}
\expandafter\ifx\csname natexlab\endcsname\relax\def\natexlab#1{#1}\fi
\expandafter\ifx\csname bibnamefont\endcsname\relax
  \def\bibnamefont#1{#1}\fi
\expandafter\ifx\csname bibfnamefont\endcsname\relax
  \def\bibfnamefont#1{#1}\fi
\expandafter\ifx\csname citenamefont\endcsname\relax
  \def\citenamefont#1{#1}\fi
\expandafter\ifx\csname url\endcsname\relax
  \def\url#1{\texttt{#1}}\fi
\expandafter\ifx\csname urlprefix\endcsname\relax\def\urlprefix{URL }\fi
\providecommand{\bibinfo}[2]{#2}
\providecommand{\eprint}[2][]{\url{#2}}

\bibitem[{\citenamefont{Scott}(1994)}]{scott94}
\bibinfo{author}{\bibfnamefont{S.~K.} \bibnamefont{Scott}},
  \emph{\bibinfo{title}{Oscillations, waves, and chaos in chemical kinetics}}
  (\bibinfo{publisher}{Oxford University Press}, \bibinfo{year}{1994}), ISBN
  \bibinfo{isbn}{0-19-855844-9}.

\bibitem[{\citenamefont{Fisher}(1937)}]{fisher37}
\bibinfo{author}{\bibfnamefont{R.~A.} \bibnamefont{Fisher}},
  \bibinfo{journal}{Annals of Eugenics} \textbf{\bibinfo{volume}{7}},
  \bibinfo{pages}{355} (\bibinfo{year}{1937}).

\bibitem[{\citenamefont{Kolmogorov et~al.}(1937)\citenamefont{Kolmogorov,
  Petrovsky, and Piscounoff}}]{kolmogorov37}
\bibinfo{author}{\bibfnamefont{A.}~\bibnamefont{Kolmogorov}},
  \bibinfo{author}{\bibfnamefont{I.}~\bibnamefont{Petrovsky}},
  \bibnamefont{and}
  \bibinfo{author}{\bibfnamefont{N.}~\bibnamefont{Piscounoff}},
  \bibinfo{journal}{Bull. Univ. Moscow, Ser. Int. A}
  \textbf{\bibinfo{volume}{1}} (\bibinfo{year}{1937}).

\bibitem[{\citenamefont{Zeldovich and Frank-Kamenetskii}(1938)}]{zeldovich38}
\bibinfo{author}{\bibfnamefont{Y.}~\bibnamefont{Zeldovich}} \bibnamefont{and}
  \bibinfo{author}{\bibfnamefont{D.}~\bibnamefont{Frank-Kamenetskii}},
  \bibinfo{journal}{Zh. Fiz. Khim} \textbf{\bibinfo{volume}{12}},
  \bibinfo{pages}{100} (\bibinfo{year}{1938}).

\bibitem[{\citenamefont{Audoly et~al.}(2000)\citenamefont{Audoly, Berestycki,
  and Pomeau}}]{audoly00}
\bibinfo{author}{\bibfnamefont{B.}~\bibnamefont{Audoly}},
  \bibinfo{author}{\bibfnamefont{H.}~\bibnamefont{Berestycki}},
  \bibnamefont{and} \bibinfo{author}{\bibfnamefont{Y.}~\bibnamefont{Pomeau}},
  \bibinfo{journal}{CR Acad. Sci. Paris, Ser. II B}
  \textbf{\bibinfo{volume}{328}}, \bibinfo{pages}{255} (\bibinfo{year}{2000}).

\bibitem[{\citenamefont{Edwards}(2002)}]{edwards02}
\bibinfo{author}{\bibfnamefont{B.~F.} \bibnamefont{Edwards}},
  \bibinfo{journal}{Phys. Rev. Lett.} \textbf{\bibinfo{volume}{89}},
  \bibinfo{pages}{104501} (\bibinfo{year}{2002}).

\bibitem[{\citenamefont{Edwards}(2006)}]{edwards06}
\bibinfo{author}{\bibfnamefont{B.~F.} \bibnamefont{Edwards}},
  \bibinfo{journal}{Chaos: An Interdisciplinary Journal of Nonlinear Science}
  \textbf{\bibinfo{volume}{16}}, \bibinfo{eid}{043106}
  (pages~\bibinfo{numpages}{8}) (\bibinfo{year}{2006}),
  \urlprefix\url{http://link.aip.org/link/?CHA/16/043106/1}.

\bibitem[{\citenamefont{Leconte et~al.}(2003)\citenamefont{Leconte, Martin,
  Rakotomalala, and Salin}}]{leconte03}
\bibinfo{author}{\bibfnamefont{M.}~\bibnamefont{Leconte}},
  \bibinfo{author}{\bibfnamefont{J.}~\bibnamefont{Martin}},
  \bibinfo{author}{\bibfnamefont{N.}~\bibnamefont{Rakotomalala}},
  \bibnamefont{and} \bibinfo{author}{\bibfnamefont{D.}~\bibnamefont{Salin}},
  \bibinfo{journal}{Phys. Rev. Lett.} \textbf{\bibinfo{volume}{90}},
  \bibinfo{pages}{128302} (\bibinfo{year}{2003}).

\bibitem[{\citenamefont{Leconte et~al.}(2004)\citenamefont{Leconte, Martin,
  Rakotomalala, and Salin}}]{leconte04}
\bibinfo{author}{\bibfnamefont{M.}~\bibnamefont{Leconte}},
  \bibinfo{author}{\bibfnamefont{J.}~\bibnamefont{Martin}},
  \bibinfo{author}{\bibfnamefont{N.}~\bibnamefont{Rakotomalala}},
  \bibnamefont{and} \bibinfo{author}{\bibfnamefont{D.}~\bibnamefont{Salin}},
  \bibinfo{journal}{J. Chem. Phys.} \textbf{\bibinfo{volume}{120}},
  \bibinfo{pages}{7314} (\bibinfo{year}{2004}).

\bibitem[{\citenamefont{Vasquez}(2007)}]{vasquez07}
\bibinfo{author}{\bibfnamefont{D.~A.} \bibnamefont{Vasquez}},
  \bibinfo{journal}{Phys. Rev. E} \textbf{\bibinfo{volume}{76}},
  \bibinfo{pages}{056308} (\bibinfo{year}{2007}).

\bibitem[{\citenamefont{Leconte et~al.}(2008)\citenamefont{Leconte, Jarrige,
  Martin, Rakotomalala, Salin, and Talon}}]{leconte08}
\bibinfo{author}{\bibfnamefont{M.}~\bibnamefont{Leconte}},
  \bibinfo{author}{\bibfnamefont{N.}~\bibnamefont{Jarrige}},
  \bibinfo{author}{\bibfnamefont{J.}~\bibnamefont{Martin}},
  \bibinfo{author}{\bibfnamefont{N.}~\bibnamefont{Rakotomalala}},
  \bibinfo{author}{\bibfnamefont{D.}~\bibnamefont{Salin}}, \bibnamefont{and}
  \bibinfo{author}{\bibfnamefont{L.}~\bibnamefont{Talon}},
  \bibinfo{journal}{Phys. Fluids} \textbf{\bibinfo{volume}{20}},
  \bibinfo{pages}{057102} (\bibinfo{year}{2008}).

\bibitem[{\citenamefont{Schwartz and Solomon}(2008)}]{schwartz08}
\bibinfo{author}{\bibfnamefont{M.~E.} \bibnamefont{Schwartz}} \bibnamefont{and}
  \bibinfo{author}{\bibfnamefont{T.~H.} \bibnamefont{Solomon}},
  \bibinfo{journal}{Phys. Rev. Lett.} \textbf{\bibinfo{volume}{100}},
  \bibinfo{pages}{028302} (\bibinfo{year}{2008}).

\bibitem[{\citenamefont{Mitchell and Mahoney}(2012)}]{mitchell12}
\bibinfo{author}{\bibfnamefont{K.~A.} \bibnamefont{Mitchell}} \bibnamefont{and}
  \bibinfo{author}{\bibfnamefont{J.~R.} \bibnamefont{Mahoney}},
  \bibinfo{journal}{Chaos} \textbf{\bibinfo{volume}{22}}, \bibinfo{eid}{037104}
  (\bibinfo{year}{2012}),
  \urlprefix\url{http://scitation.aip.org/content/aip/journal/chaos/22/3/10.1063/1.4746039}.

\bibitem[{\citenamefont{Bargteil and Solomon}(2012)}]{bargteil12}
\bibinfo{author}{\bibfnamefont{D.}~\bibnamefont{Bargteil}} \bibnamefont{and}
  \bibinfo{author}{\bibfnamefont{T.}~\bibnamefont{Solomon}},
  \bibinfo{journal}{Chaos: An Interdisciplinary Journal of Nonlinear Science}
  \textbf{\bibinfo{volume}{22}}, \bibinfo{pages}{037103}
  (\bibinfo{year}{2012}).

\bibitem[{\citenamefont{Megson et~al.}(2015)\citenamefont{Megson, Najarian,
  Lilienthal, and Solomon}}]{megson15}
\bibinfo{author}{\bibfnamefont{P.~W.} \bibnamefont{Megson}},
  \bibinfo{author}{\bibfnamefont{M.~L.} \bibnamefont{Najarian}},
  \bibinfo{author}{\bibfnamefont{K.~E.} \bibnamefont{Lilienthal}},
  \bibnamefont{and} \bibinfo{author}{\bibfnamefont{T.~H.}
  \bibnamefont{Solomon}}, \bibinfo{journal}{Physics of Fluids}
  \textbf{\bibinfo{volume}{27}}, \bibinfo{eid}{023601} (\bibinfo{year}{2015}),
  \urlprefix\url{http://scitation.aip.org/content/aip/journal/pof2/27/2/10.1063/1.4913380}.

\bibitem[{\citenamefont{Mahoney et~al.}(2015)\citenamefont{Mahoney, Li, Boyer,
  Solomon, and Mitchell}}]{mahoney15}
\bibinfo{author}{\bibfnamefont{J.~R.} \bibnamefont{Mahoney}},
  \bibinfo{author}{\bibfnamefont{J.}~\bibnamefont{Li}},
  \bibinfo{author}{\bibfnamefont{C.}~\bibnamefont{Boyer}},
  \bibinfo{author}{\bibfnamefont{T.}~\bibnamefont{Solomon}}, \bibnamefont{and}
  \bibinfo{author}{\bibfnamefont{K.~A.} \bibnamefont{Mitchell}},
  \bibinfo{journal}{Physical Review E} \textbf{\bibinfo{volume}{92}},
  \bibinfo{pages}{063005} (\bibinfo{year}{2015}).

\bibitem[{\citenamefont{Atis et~al.}(2012)\citenamefont{Atis, Saha, Auradou,
  Martin, Rakotomalala, Talon, and Salin}}]{atis12}
\bibinfo{author}{\bibfnamefont{S.}~\bibnamefont{Atis}},
  \bibinfo{author}{\bibfnamefont{S.}~\bibnamefont{Saha}},
  \bibinfo{author}{\bibfnamefont{H.}~\bibnamefont{Auradou}},
  \bibinfo{author}{\bibfnamefont{J.}~\bibnamefont{Martin}},
  \bibinfo{author}{\bibfnamefont{N.}~\bibnamefont{Rakotomalala}},
  \bibinfo{author}{\bibfnamefont{L.}~\bibnamefont{Talon}}, \bibnamefont{and}
  \bibinfo{author}{\bibfnamefont{D.}~\bibnamefont{Salin}},
  \bibinfo{journal}{Chaos: An Interdisciplinary Journal of Nonlinear Science}
  \textbf{\bibinfo{volume}{22}}, \bibinfo{eid}{037108}
  (pages~\bibinfo{numpages}{11}) (\bibinfo{year}{2012}),
  \urlprefix\url{http://link.aip.org/link/?CHA/22/037108/1}.

\bibitem[{\citenamefont{Saha et~al.}(2013)\citenamefont{Saha, Atis, Salin, and
  Talon}}]{saha13}
\bibinfo{author}{\bibfnamefont{S.}~\bibnamefont{Saha}},
  \bibinfo{author}{\bibfnamefont{S.}~\bibnamefont{Atis}},
  \bibinfo{author}{\bibfnamefont{D.}~\bibnamefont{Salin}}, \bibnamefont{and}
  \bibinfo{author}{\bibfnamefont{L.}~\bibnamefont{Talon}},
  \bibinfo{journal}{EPL} \textbf{\bibinfo{volume}{101}}, \bibinfo{pages}{38003}
  (\bibinfo{year}{2013}),
  \urlprefix\url{http://stacks.iop.org/0295-5075/101/i=3/a=38003}.

\bibitem[{\citenamefont{Mahoney et~al.}(2012)\citenamefont{Mahoney, Bargteil,
  Kingsbury, Mitchell, and Solomon}}]{mahoney12}
\bibinfo{author}{\bibfnamefont{J.}~\bibnamefont{Mahoney}},
  \bibinfo{author}{\bibfnamefont{D.}~\bibnamefont{Bargteil}},
  \bibinfo{author}{\bibfnamefont{M.}~\bibnamefont{Kingsbury}},
  \bibinfo{author}{\bibfnamefont{K.}~\bibnamefont{Mitchell}}, \bibnamefont{and}
  \bibinfo{author}{\bibfnamefont{T.}~\bibnamefont{Solomon}},
  \bibinfo{journal}{EPL (Europhysics Letters)} \textbf{\bibinfo{volume}{98}},
  \bibinfo{pages}{44005} (\bibinfo{year}{2012}).

\bibitem[{\citenamefont{Hanna et~al.}(1982)\citenamefont{Hanna, Saul, and
  Showalter}}]{hanna82}
\bibinfo{author}{\bibfnamefont{A.}~\bibnamefont{Hanna}},
  \bibinfo{author}{\bibfnamefont{A.}~\bibnamefont{Saul}}, \bibnamefont{and}
  \bibinfo{author}{\bibfnamefont{K.}~\bibnamefont{Showalter}},
  \bibinfo{journal}{J. Am. Chem. Soc.} \textbf{\bibinfo{volume}{104}},
  \bibinfo{pages}{3838} (\bibinfo{year}{1982}).

\bibitem[{\citenamefont{Yoshinaga et~al.}(2004)\citenamefont{Yoshinaga,
  Tsuschida, Toyose, Hiratsuka, and Yamaye}}]{yoshinaga04}
\bibinfo{author}{\bibfnamefont{T.}~\bibnamefont{Yoshinaga}},
  \bibinfo{author}{\bibfnamefont{M.}~\bibnamefont{Tsuschida}},
  \bibinfo{author}{\bibfnamefont{Y.}~\bibnamefont{Toyose}},
  \bibinfo{author}{\bibfnamefont{H.}~\bibnamefont{Hiratsuka}},
  \bibnamefont{and} \bibinfo{author}{\bibfnamefont{M.}~\bibnamefont{Yamaye}},
  \bibinfo{journal}{Anal. Sci.} \textbf{\bibinfo{volume}{20}},
  \bibinfo{pages}{549} (\bibinfo{year}{2004}).

\bibitem[{\citenamefont{B{\"o}ckmann and M{\"u}ller}(2000)}]{bockmann00}
\bibinfo{author}{\bibfnamefont{M.}~\bibnamefont{B{\"o}ckmann}}
  \bibnamefont{and} \bibinfo{author}{\bibfnamefont{S.~C.}
  \bibnamefont{M{\"u}ller}}, \bibinfo{journal}{Phys. Rev. Lett.}
  \textbf{\bibinfo{volume}{85}}, \bibinfo{pages}{2506} (\bibinfo{year}{2000}).

\bibitem[{\citenamefont{Hele-Shaw}(1898)}]{Heleshaw98}
\bibinfo{author}{\bibfnamefont{H.~J.~S.} \bibnamefont{Hele-Shaw}},
  \bibinfo{journal}{Nature} \textbf{\bibinfo{volume}{58}}, \bibinfo{pages}{34}
  (\bibinfo{year}{1898}).

\bibitem[{\citenamefont{Lamb}(1932)}]{lamb32}
\bibinfo{author}{\bibfnamefont{H.}~\bibnamefont{Lamb}},
  \emph{\bibinfo{title}{Hydrodynamics, 6th ed.}} (\bibinfo{publisher}{Cambridge
  University Press}, \bibinfo{year}{1932}).

\bibitem[{\citenamefont{Lee and Fung}(1969)}]{lee69}
\bibinfo{author}{\bibfnamefont{J.~S.} \bibnamefont{Lee}} \bibnamefont{and}
  \bibinfo{author}{\bibfnamefont{Y.~C.} \bibnamefont{Fung}},
  \bibinfo{journal}{Journal of Fluid Mechanics} \textbf{\bibinfo{volume}{37}},
  \bibinfo{pages}{657} (\bibinfo{year}{1969}), ISSN \bibinfo{issn}{1469-7645},
  \urlprefix\url{http://journals.cambridge.org/article_S0022112069000796}.

\bibitem[{\citenamefont{Ginzburg et~al.}(2010)\citenamefont{Ginzburg,
  {d'}Humi{`e}res, and Kuzmin}}]{ginzburg10}
\bibinfo{author}{\bibfnamefont{I.}~\bibnamefont{Ginzburg}},
  \bibinfo{author}{\bibfnamefont{D.}~\bibnamefont{{d'}Humi{`e}res}},
  \bibnamefont{and} \bibinfo{author}{\bibfnamefont{A.}~\bibnamefont{Kuzmin}},
  \bibinfo{journal}{Journal of Statistical Physics}
  \textbf{\bibinfo{volume}{139}}, \bibinfo{pages}{1090} (\bibinfo{year}{2010}),
  ISSN \bibinfo{issn}{0022-4715}, \bibinfo{note}{10.1007/s10955-010-9969-9},
  \urlprefix\url{http://dx.doi.org/10.1007/s10955-010-9969-9}.

\bibitem[{\citenamefont{Guyon et~al.}(2001)\citenamefont{Guyon, Hulin, Petit,
  and de~Gennes}}]{guyon01}
\bibinfo{author}{\bibfnamefont{E.}~\bibnamefont{Guyon}},
  \bibinfo{author}{\bibfnamefont{J.-P.} \bibnamefont{Hulin}},
  \bibinfo{author}{\bibfnamefont{L.}~\bibnamefont{Petit}}, \bibnamefont{and}
  \bibinfo{author}{\bibfnamefont{P.~G.} \bibnamefont{de~Gennes}},
  \emph{\bibinfo{title}{Hydrodynamique physique}} (\bibinfo{publisher}{EDP
  sciences Les Ulis}, \bibinfo{year}{2001}).

\end{thebibliography}
\end{document}